\documentclass[fleqn,usenatbib]{mnras}

\usepackage{newtxtext,newtxmath}

\usepackage[T1]{fontenc}
\usepackage{ae,aecompl}

\usepackage{graphicx}	
\usepackage{amsmath}	
\usepackage[utf8]{inputenc}

\title[UV suppression of ASASSN-14lp]{ASASSN-14lp: two possible solutions for the observed UV suppression}
\author[B. Barna et al.]{Barnabas Barna,$^{1,2}$\thanks{E-mail: bbarna@titan.physx.u-szeged.hu}
Talytha Pereira,$^{3}$
Stefan Taubenberger,$^{3}$
Mark Magee,$^{4}$\newauthor
Markus Kromer,$^{5,6}$
Wolfgang Kerzendorf,$^{7,8}$
Christian Vogl,$^{3,9,14}$
Marc E. Williamson,$^{10}$ \newauthor
Andreas Flörs,$^{3,9,11}$
Ulrich M. Noebauer,$^{3}$
Ryan J. Foley,$^{12}$
Michele Sasdelli$^{13}$ \newauthor
and Wolfgang Hillebrandt$^{3}$
\\
$^{1}$Astronomical Institute, Academy of Sciences of the Czech Republic, Bo\u{c}ni II 1401, 141 31 Prague, Czech Republic\\
$^{2}$Physics Institute, University of Szeged, D\'{o}m t\'{e}r 9, Szeged, 6723, Hungary\\
$^{3}$Max-Planck-Institut für Astrophysik, Karl-Schwarzschild-Str. 1, 85748 Garching, Germany\\
$^{4}$School of Physics, Trinity College Dublin, University of Dublin, Dublin 2, Ireland\\
$^{5}$Zentrum für Astronomie der Universität Heidelberg, Institut für Theoretische Astrophysik, Philosophenweg 12, D-69120 Heidelberg, Germany\\
$^{6}$Heidelberger Institut für Theoretische Studien, Schloss-Wolfsbrunnenweg 35, D-69118 Heidelberg, Germany\\
$^{7}$Department of Physics and Astronomy, Michigan State University, East Lansing, MI 48824, USA\\
$^{8}$Department of Computational Mathematics, Science, and Engineering, Michigan State University, East Lansing, MI 48824, USA\\
$^{9}$Physik Department, Technische Universität München, James-Franck-Str. 1, 85741 Garching, Germany\\
$^{10}$New York University, New York, NY 10003, USA\\
$^{11}$European Southern Observatory, Karl-Schwarzschild-Straße 2, D-85748 Garching bei München, Germany\\
$^{12}$Department of Astronomy and Astrophysics, University of California, Santa Cruz, CA 95064, USA\\
$^{13}$Australian Institute for Machine Learning, University of Adelaide, Adelaide, Australia\\
$^{14}$Exzellenzcluster ORIGINS, Boltzmannstr. 2, 85748 Garching, Germany
}

\date{Accepted XXX. Received YYY; in original form ZZZ}

\pubyear{2019}

\begin{document}
\label{firstpage}
\pagerange{\pageref{firstpage}--\pageref{lastpage}}
\maketitle

\begin{abstract}
We test the adequacy of ultraviolet (UV) spectra for characterizing the outer structure of Type Ia supernova (SN) ejecta. For this purpose, we perform spectroscopic analysis for ASASSN-14lp, a normal SN Ia showing low continuum in the mid-UV regime. To explain the strong UV suppression, two possible origins have been investigated by mapping the chemical profiles over a significant part of their ejecta. We fit the spectral time series with mid-UV coverage obtained before and around maximum light by HST, supplemented with ground-based optical observations for the earliest epochs. The synthetic spectra are calculated with the one dimensional MC radiative-transfer code TARDIS from self-consistent ejecta models. Among several physical parameters, we constrain the abundance profiles of nine chemical elements.
We find that a distribution of $^{56}$Ni (and other iron-group elements) that extends toward the highest velocities reproduces the observed UV flux well. The presence of radioactive material in the outer layers of the ejecta, if confirmed, implies strong constraints on the possible explosion scenarios. We investigate the impact of the inferred $^{56}$Ni distribution on the early light curves with the radiative transfer code TURTLS, and confront the results with the observed light curves of ASASSN-14lp. The inferred abundances are not in conflict with the observed photometry. 
We also test whether the UV suppression can be reproduced if the radiation at the photosphere is significantly lower in the UV regime than the pure Planck function. In this case, solar metallicity might be sufficient enough at the highest velocities to reproduce the UV suppression.

\end{abstract}




\section{Introduction}
\label{sec:introduction}

Type Ia SNe (SNe~Ia) originate from thermonuclear explosions \citep{Hoyle60} of white dwarfs (WDs). The released energy completely disrupts the WD in most cases and the ejecta starts a homologous expansion. During the first months, the decay of the produced $^{56}$Ni powers the light curve (LC), which reaches its peak in three weeks after the explosion. The peak luminosity strongly correlates with the decline-rate, allowing SNe Ia to be used as standardizable candles, measuring distances even on cosmological scales \citep{Riess98,Perlmutter99}.

However, SNe Ia show a strong variation in the rise time, thus, likely in the distribution of $^{56}$Ni \citep{Woosley07,Piro13,Piro14,Piro16,Magee20}. \cite{Firth15} studied a sample of 18 normal SNe Ia and found rise times of bolometric LCs ranging from 15.98 to 24.7 days with a mean value of 18.98 days. 
The variations in location and abundance of $^{56}$Ni could originate from the different explosions of the individual objects.

Although SNe Ia are among the most-studied astrophysical objects, the nature of their progenitor system and their explosion mechanism are still under debate. In the classic single-degenerate (SD) scenario \citep{Whelan73}, the WD accretes material from a main-sequence or a giant star and the ignition happens in the core of the WD. However, a thermonuclear explosion could also arise from a system with two WDs (double-degenerate or DD scenario), where the main component either tidally disrupts the secondary component and accretes its material or the two WDs merge violently \citep{Iben84,Webbink84,Pakmor11}. 

There are more distinct modes of how the explosion could take place after the ignition of a WD, distinguished mainly by the propagation of the fusion flame. A deflagration propagates through the conduction of heat, which ignites the fuel ahead of the burning front \citep{Timmes92}. This process can be accelerated by thermodynamic instabilities, which wrinkle the burning front and increase its surface. Since the propagation of a deflagration is subsonic, the WD can react by expansion, and much of the burning therefore takes place at relatively low densities in the outer layers, leaving behind intermediate-mass elements (IMEs) mostly. The outermost layers, which are not reached by the fusion flame, still consist of the original C/O matter of the WD. However, hydrodynamic models, where only deflagration takes place, predict strong mixing in the ejecta due to Rayleigh-Taylor and Kelvin-Helmholtz instabilities. Therefore, these so-called pure deflagration models are expected to have a nearly uniform chemical structure, producing $^{56}$Ni mass fraction of 0.2-0.5 even at the top of the ejecta. Moreover, these explosions are not energetic enough to accelerate significant mass to high velocities, thus, no line formation is expected over 15\,000 km s$^{-1}$, and pure deflagration may not able to fully unbound the WD and leave behind a bound remnant \citep{Fink14}.

Pure deflagrations (even in extreme cases) produce less than 0.4 M$_\odot$ $^{56}$Ni, not enough for 'normally bright' SNe Ia, and do not give 'normally' bright supernovae but predict 'faint' explosions instead. The supersonic propagation of the burning front called detonation compresses and ignites the material in front of the fusion front \citep{Gamezo99}. The energy released from the combustion maintains the propagation of the shock wave. Leaving the WD ahead of the shock no time to expand, a detonation in a Chandrasekhar-mass WD results in an ejecta almost completely burned to iron-group elements (IGEs) even at high velocities. Assuming a Chandrasekhar-mass ($M_\rmn{Ch}$) WD as the progenitor, neither the deflagration nor the detonation scenario alone can fully explain the observed spectral features of normal SNe Ia  \citep{Ropke17}. Thus, a deflagration-to-detonation transition (DDT) is postulated to get 'normal' SNe Ia from near-Chandrasekhar WDs, however, whether or not such DDT can happen is still unclear \citep{Ropke17}. \cite{Khokhlov91} assumed a spontaneous transition from the initial deflagration to a detonation forming an IGE dominated inner and unburnt material dominated outer region, separated by an IME dominated intermediate region. The location of the three characteristic domains may vary depending on the luminosity of the object \citep{Hoflich02}. In the case of a typical 'normal' SN Ia like SN 2011fe, the separation of IME- and C/O dominated regions is expected to appear at $\sim$14\,000 km s$^{-1}$ \citep{Mazzali14}.

As an alternative, sub-Chandrasekhar WDs could also be progenitors in the so-called double detonation models \citep{Woosley94,Livne95,Fink07}. In this scenario, the first detonation occurs in the outer He-rich shell accreted by the WD from its companion star. This triggers a secondary detonation in the core of the WD, which fully disrupts the progenitor. The double-detonation scenario may result in a wide range of explosion energies due to the diverse masses of the progenitor and He-shell. As a common feature, hydrodynamic simulations of the DD scenario \citep[see e.g.][]{Fink10,Kromer10} predict an IGE dominated layer above the IME region compared to the DDT explosions, although the expected location and mass of the outer $^{56}$Ni synthesized in the first detonation strongly depends on the initial condition of the He-layer. The significant amount IGEs at higher (> 10\,000 km s$^{-1}$) velocities may be a unique property of the DD scenario, which also leaves its imprint in the observables.

\begin{table*}
	\centering
	\caption{Log of the spectra of our sample; the phases are given with respect to B-band maximum. The times since explosion ($t_\rmn{exp}$) were fitting parameters for our TARDIS models (see in Sec. \ref{sec:fitting}) within $\pm$1.5 days of their estimated values in the referred papers.}
    \label{tab:log14lp}
    \begin{tabular}{cccccc}
		\hline
		MJD & Phase [days] & Telescope / Instrument & Wavelength [\r{A}] & Paper\\
        \hline
        \multicolumn{5}{c}{ASASSN-14lp}\\
		\hline
        57001.55 & -13.8 & MDM2.4m/ModSpec & 4200 -- 7400 & \cite{Shappee16}\\
        57003.52 & $-11.8$ & APO3.5m/DIS & 3500 -- 9600 & \cite{Shappee16}\\
        57006.13 & $-9.2$ & HST/STIS & 1600 -- 10200 & \cite{Foley16}\\
        57008.89 & $-6.4$ & HST/STIS & 1600 -- 10200 & \cite{Foley16}\\
        57010.81 & $-4.5$ & HST/STIS & 1600 -- 10200 & \cite{Foley16}\\
        57013.86 & $-1.5$ & HST/STIS & 1600 -- 10200 & \cite{Foley16}\\
        57015.65 & $+0.3$ & HST/STIS & 1600 -- 10200 & \cite{Foley16}\\
        57017.64 & $+2.3$ & HST/STIS & 1600 -- 10200 & \cite{Foley16}\\
        57020.43 & $+5.1$ & HST/STIS & 1600 -- 10200 & \cite{Foley16}\\
        57023.41 & $+8.1$ & HST/STIS & 1600 -- 10200 & \cite{Foley16}\\
        \hline
	\end{tabular}
\end{table*}

An additional observational property of SNe Ia, which is strongly linked to the abundances of IGEs and, thereby, to $^{56}$Ni, is the UV-diversity reported in several papers \citep[e.g.][]{Brown15,Foley16}. Below $\sim$3000 \r{A}, the opacity is dominated by the overlapping lines of IGEs, mainly iron and nickel \citep{Baron96,Pinto00}. A UV photon is repeatedly absorbed and emitted by IGE atoms in the SN atmosphere, until a scattering changes its wavelength redward and it can escape from the SN. This fluorescence process redistributes flux from the UV (100-4000 \r{A}) to the optical regime resulting in flux suppression at shorter wavelengths. However, the strength of this suppression is highly diverse especially in the mid-UV (2000-3000 \r{A}) as it was shown both by photometric \citep{Brown10} and spectroscopic \citep{Foley16} studies.

The exact relation between the UV suppression and other SN properties is still unclear, although it was shown by various theoretical studies that the UV flux is primarily sensitive to the metallicity of the SN \citep{Lentz00,Sauer08,Walker12}. The increased metal content could originate either directly from the chemical properties of the progenitor WD \citep{Hoflich98} or be produced in the nucleosynthesis of the explosion. Since the UV suppression is observed even at the pre-maximum epochs, it should be sensitive to the composition of the outer ejecta (i.e. at velocities above $\sim$12\,000 km s$^{-1}$).

As the different explosion scenarios result in a distinct distribution of the chemical elements, observational properties like UV suppression may eventually provide constraints on the explosion mechanism of the studied SNe. We aim to extract the information contained in UV spectra by performing an abundance tomography analysis \citep[see e.g.][]{Stehle05}. The technique allows us to reconstruct the structure of the outer ejecta. Thus, if the above mentioned assumptions are correct, we hope to identify the origin of the UV-suppression, which may eventually provide constraints on the explosion mechanism of the studied SNe.

The paper is structured in the following order. In Section \ref{sec:sn2014lp}, we describe the target of this study, ASASSN-14p, a normal Type Ia SN showing strong UV suppression, and we give an overview of our data sample. In Section \ref{sec:fitting}, we describe our spectral fitting procedure to perform an abundance tomography analysis. In Section \ref{sec:results}, we present the results of our spectral modelling, focusing on the chemical structure of the outer ejecta, which may explain the low flux level in the mid-UV. The inferred $^{56}$Ni distribution is also tested against the light curves. We investigate the effect of the alternation of the lower-boundary radiation field in our modelling approach, as another possible origin for the observed UV-suppression. We also performed abundance tomography on SN 2013dy, another normal SN Ia showing UV suppression. The results, which support those presented about ASASSN-14lp, are described in the \ref{appendix:sn2013dy}. Finally, we summarize our main findings in Section \ref{sec:conclusions}.

\section{ASASSN-14lp}
\label{sec:sn2014lp}

\begin{figure}
	\includegraphics[width=\columnwidth]{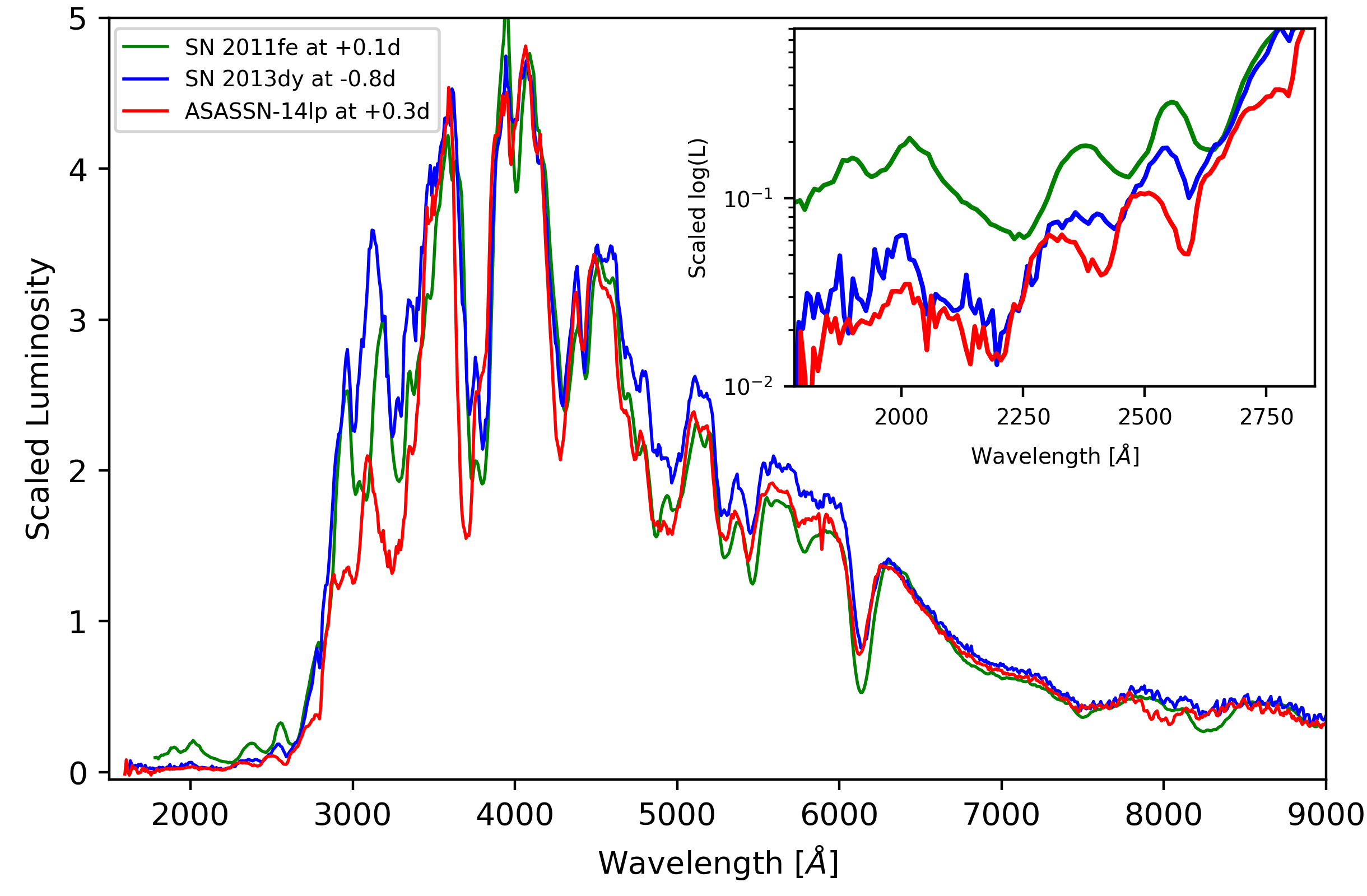}
    \caption{Comparison of the near-maximum spectrum of ASASSN-14lp with that of SNe 2011fe \citep[lacking of strong UV suppression][]{Mazzali14} and 2013dy \citep[also showing UV suppression][]{Zheng13}. The spectra are corrected reddening and redshift.}
    \label{fig:sne_ia_comp}
\end{figure}

Our study targets a normal SN Ia showing strong flux-suppression at the mid-UV wavelengths at early epochs (see in Fig. \ref{fig:sne_ia_comp}). These constraints limit the sample to only a few objects, which were the targets of HST spectroscopic follow-up. Moreover, the abundance tomography analysis requires spectral time series with a cadence of $2-3$ days, starting as early as possible after the explosion. Meeting all these requirements, ASASSN-14lp was chosen as the target of our analysis. ASASSN-14lp has been the subject of a recent study by \cite{Chen19}, where the authors used artificial intelligence assisted inversion (AIAI) for fitting spectra of SNe Ia with TARDIS. Note that despite the same modelling environment, their fitting method is different, and the fitting is limited to individual spectra, instead of performing a full abundance tomography. 

ASASSN-14lp was discovered by the All-Sky Automated Survey for SuperNovae (ASASSN) on MJD 57\,000.61, close to the center of NGC 4666 ($z=0.0051$). The inferred date of first-light from the power-law fit of the light curve is MJD $56\,998.38 \pm 0.1$ \citep{Shappee16}, while the estimated date of explosion is MJD $56\,997.88 \pm 0.48$ \citep{Magee20} based on modelling the distribution of $^{56}$Ni by TURTLS.

Its location results in a strong extragalactic extinction, which has been estimated with SNooPy to be $E(B-V)_\rmn{host}=0.33$ mag by \cite{Shappee16} based on the best-observed calibration of \cite{Folatelli10}. Note that the latter reference reported $R_\rmn{V}=1.74$ for their sample of SNe Ia corresponding to a steeper extinction curve than the usually adopted $R_\rmn{V}\,=\,3.1$. However, the reddening corrected spectra resulting from these values showed an extremely blue continuum, which we could not reproduce in TARDIS with conventional choices of the physical parameters. Keeping $R_V\,=\,1.74$, we reduced the host reddening to $E(B-V)_\rmn{host}\,=\,0.25$ mag, which is still within the 2$\sigma$ range of the SNooPy fit of \cite{Shappee16}. 

ASASSN14-lp was also the subject of spectroscopic follow-up by the HST STIS, providing nine spectral epochs in the first five weeks of the SN. Additional optical spectroscopy was done with the 2.4-m Hiltner telescope of the MDM Observatory and the 1.5-m Tillinghast telescope of the F. L. Whipple Observatory. These spectra supplement the HST spectral coverage, resulting in a time resolution of two days at epochs between -10 and +17 days with respect to the time of B-band maximum (MJD 57\,015.3). The full list of spectroscopic observations used in this study can be found in Table \ref{tab:log14lp}.

\section{Spectral analysis}
\label{sec:fitting}

We apply the technique called abundance tomography \citep{Stehle05}, which aims at constraining the chemical distribution of the ejecta via spectral fitting. As the supernova expands, the ejecta becomes optically thinner with time and the photosphere recedes into deeper layers. Thus, spectroscopic observations at different phases can probe different layers. Fitting the spectral time series of an SN with the same self-consistent model can map the physical and chemical properties of the ejecta. The abundance tomography technique was first performed by \cite{Stehle05} for SN 2002bo. Well-observed SNe Ia, such as SN 2011fe \citep{Mazzali14} and SN 2014J \citep{Ashall14} were also subjects of abundance tomography studies, just like several Type Iax SNe \citep[e.g.][]{Barna18}.

For performing abundance tomography, we use the one dimensional Monte Carlo radiative transfer code TARDIS \citep{Kerzendorf14}. The code assumes a sharp photosphere emitting blackbody radiation. The part of the model ejecta above the photosphere (hereafter referred to as atmosphere) is divided into radial layers with densities and chemical abundances defined by the user. Photon packets are injected at the bottom layer of the computation volume, i.e. the photosphere, and their propagation through the model ejecta is followed by the code. The escaping photon packets are summarized according to their wavelengths and a synthetic spectrum is built. 

The approach of TARDIS offers several improvements from the simple LTE assumptions. We followed the same method as in any other study that used TARDIS for spectral fitting \citep[see e.g.][]{Magee16, Boyle17, Barna18}. As an example, the ionization rates used here are estimated following a non-LTE (NLTE) mode, the so-called nebular approximation, which significantly deviates from the LTE method by accounting for fraction of recombinations returning directly to the ground-state. The excitation rates are calculated according to the dilute radiation field approximation which is also not purely thermal. The summary of TARDIS numerical parameters and modes adopted in this study are listed in Tab. \ref{tab:tardis}. The limitation of the simulation background, as well as the detailed description of the NLTE methods, are presented in the original TARDIS paper \citep{Kerzendorf14}.

\cite{Shen21} presented a comparative test between LTE and fully NLTE on the synthetic spectra of WD detonations, which show UV-suppression similar to that observed in the ASASSN-14lp spectra. Before and around the maximum-light, the LTE and NLTE spectra were found almost identical in the optical and barely different in the near-UV regime. The flux suppression below 3000 \r{A} appears in both spectra similarly. However, the synthetic spectra from LTE and non-LTE calculations start to deviate for later epochs. This is mainly the effect of the hotter line-forming regions in NLTE providing a higher ratio of doubly to single-ionized elements, and thus, a weaker line blanketing by the Fe II lines. At two weeks after maximum light, the discrepancies grow strong at the UV wavelengths as well, which sets a strong time-limitation on our modeling attempts for ASASSN-14lp. This result also confirms the previous tests by \citep{Kerzendorf14} and allows us to use the assumptions of TARDIS to study the shorten wavelengths before +10 days after B-band maximum. Note that these tests do not investigate the effect of the assumed photosphere, which is the subject of Sec. \ref{sec:photosphere}.

\begin{table}
	\centering
	\caption{Summary of the used TARDIS modes and simulation settings.}
    \label{tab:tardis}
    \begin{tabular}{lcc}
		\hline
		\hline
		Settings &  & Value \\
        \hline
        Ionization mode&  & nebular\\
        Excitation mode&  & dilute-LTE\\
        Radiative rate mode &   & dilute-blackbody\\
        Line excitation mode &   & macroatom\\
        Number of iterations &   & 20\\
        Number of photon packets &   & 300\,000\\
        \hline
        \hline
	\end{tabular}
\end{table}

In this study, we mainly follow the fitting strategy described in \cite{Barna17, Barna18} with slight modifications. Note that because of the diverse nature of the spectra we cannot measure the goodness of the fits with a direct, quantitative technique, only a fit-by-eye evaluation is possible for the spectral fits as it was applied in all the previous studies \citep[see e.g.][]{Mazzali08,Ashall14,Ashall16,Magee16}. For the model structure, we adopted a pure exponential function as the density profile:

\begin{equation}
\rho (v,t_{\rmn{exp}}) = \rho_0 \cdot \left(\frac{t_{\rmn{exp}}}{t_0}\right)^{-3} \cdot \exp\left(-\frac{v}{v_0}\right),
\end{equation}

\begin{figure*}
	\includegraphics[width=18.0cm]{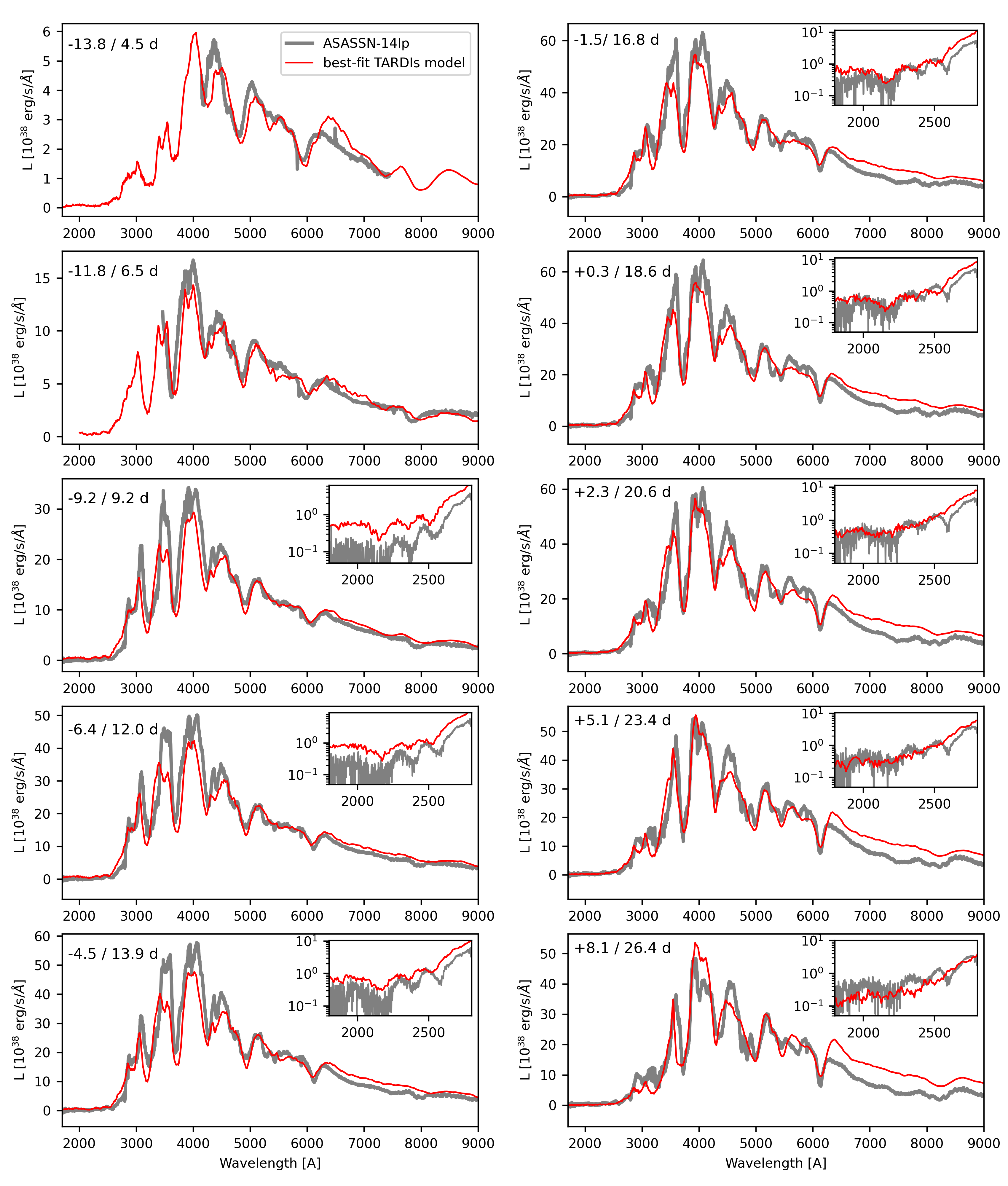}
    \caption{The studied spectral series of ASASSN-14lp (grey) and the corresponding best-fit TARDIS synthetic spectra (red).}
    \label{fig:spectra_14lp}
\end{figure*}

where $\rho_\rmn{0}$ is the inner density at the reference time chosen of $t_\rmn{0} = 100$ s, $t_\rmn{exp}$ is the time since the explosion, $v$ is the velocity coordinate and $v_\rmn{0}$ is the reference velocity, which defines the slope of the density profile. This simplified model structure allows to constrain the mass distribution in the ejecta with only two fitting parameters ($\rho_\rmn{0}$ and $v_0$). 

The model ejecta was split into multiple layers, where the chemical abundances were fitting parameters to constrain the stratification of the chemical elements. These fitting layers were defined independently from the actual photospheric velocity and set equally in velocity space. The velocity step of the boundaries was 2\,000 km s$^{-1}$ above 16\,000 km s$^{-1}$ and 1\,000 km s$^{-1}$ for the regions below that. We fit the mass fractions of seven stable elements, namely C, O, Mg, Si, S, Ca, Fe, which include the contribution of all the corresponding stable isotopes. The mass fraction of two radioactive isotopes, $^{52}$Fe and $^{56}$Ni were also subject to fitting. The decay of radioactive material, as well as the contribution of their daughter isotopes ($^{52}$Mn, $^{52}$Cr for $^{52}$Fe and $^{56}$Co, $^{56}$Fe for $^{56}$Ni) were also taken into account. Since the solar metallicity considered for every other chemical element heavier than He had no significant impact on the spectra, these mass fractions were set to zero.

When fitting each spectrum, we constructed the actual computation volume for the radiative transfer calculation based on the model structure described in the previous paragraph. The inner boundary of the computational volume is emitting pure blackbody radiation, thus, it is referred hereafter as photosphere, and its location as photospheric velocity ($v_\rmn{phot}$). The outer limit (i.e. the surface of the SN ejecta) of the volume is set to a high velocity, where the local density is sufficiently low for not having an impact on the computed spectra. According to our preliminary tests, we chose 35\,000 km s$^{-1}$ as the upper boundary in this study. The computational volume is divided into twenty radial layers, where layer boundaries are estimated according to equal mass regions. The mass fractions and the density value are sampled from the model structure and assumed to be uniform within a layer. The synthetic spectrum estimated by TARDIS based on the computational volume is compared to the observed spectrum and we modify the model structure to achieve a better match for the next TARDIS calculation. This iterative fitting allows us to constrain the physical and chemical properties that have a major impact on the output spectrum.

\begin{figure}
	\includegraphics[width=\columnwidth]{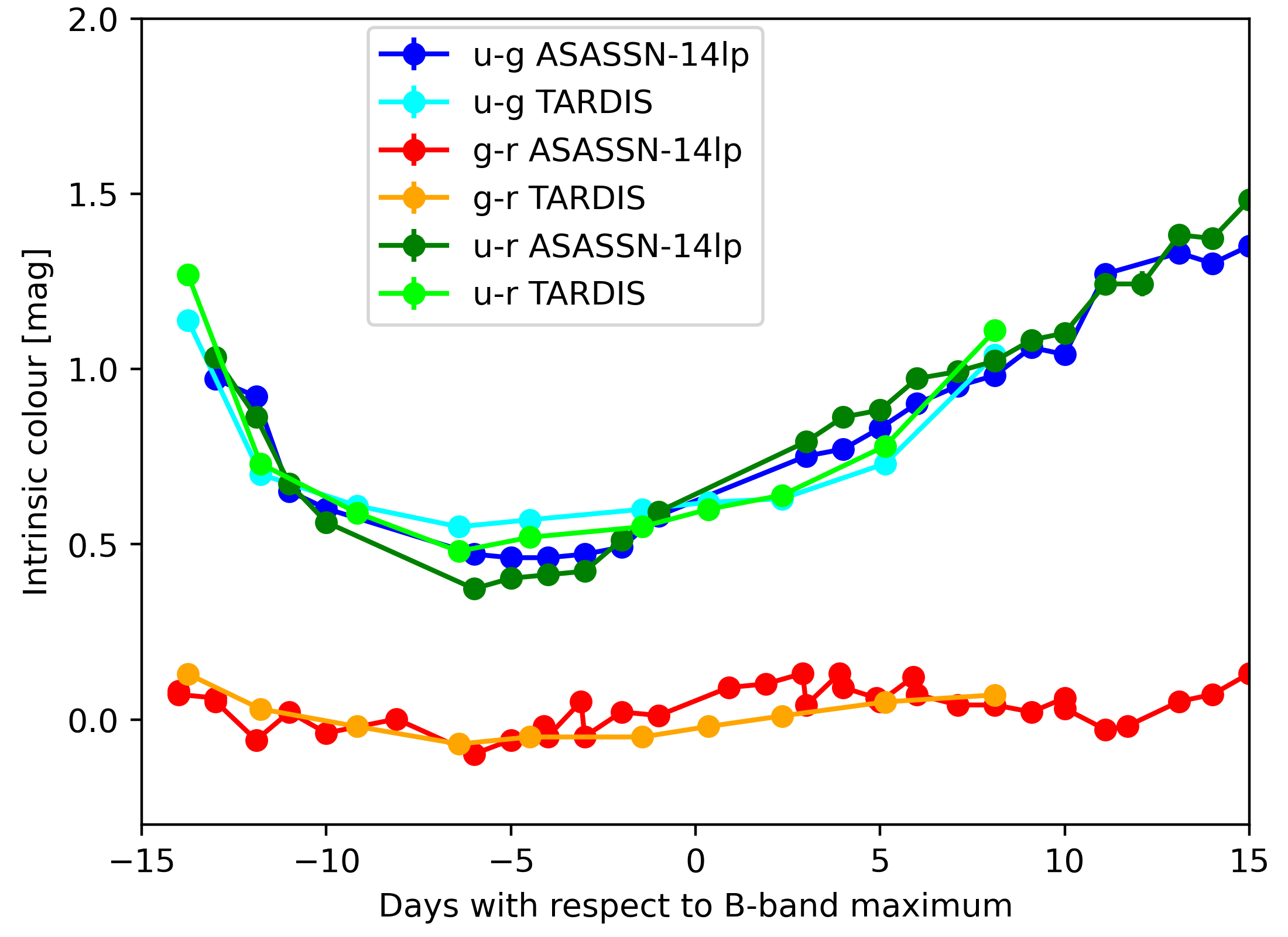}
    \caption{The intrinsic colours estimated from the \textit{ugr} photometry of ASASSN-14lp and the corresponding colours estimated from best-fit TARDIS synthetic spectra.}
    \label{fig:colors}
\end{figure}

In addition to ASASSN-14lp, we also performed abundance tomography analysis on SN 2013dy, a Type Ia SN, which also meets the criteria described in Sec. \ref{sec:sn2014lp}, although the observed UV suppression is slightly weaker. The spectral modelling of SN 2013dy and the main conclusions are very similar to the case of ASASSN-14lp, thus, we focus only on the latter object hereafter. The description of SN 2013dy, the data sets, and the abundance tomography results are included in the \ref{appendix:sn2013dy}.  

\section{Results and discussion}
\label{sec:results}

\subsection{Abundance tomography of ASASSN-14lp}

The final synthetic spectra of ASASSN-14lp can be seen in Fig. \ref{fig:spectra_14lp}. Note that the whole parameter space cannot be fully explored manually, thus, we cannot rule out a different combination of physical and chemical properties, which would provide a similarly good match with the observed spectra. However, considering that the main spectral features of each epoch are reproduced with special attention to the UV suppression (see Sec. \ref{sec:IGE_impact}), and the intrinsic colours estimated from the synthetic spectra show a good match with the those of ASASSN-14lp (see in Fig. \ref{fig:colors}), 
our results provide a feasible solution. For simplicity, we refer to the inferred synthetic ejecta structures as best-fit models hereafter.

\begin{figure*}
	\includegraphics[width=16cm]{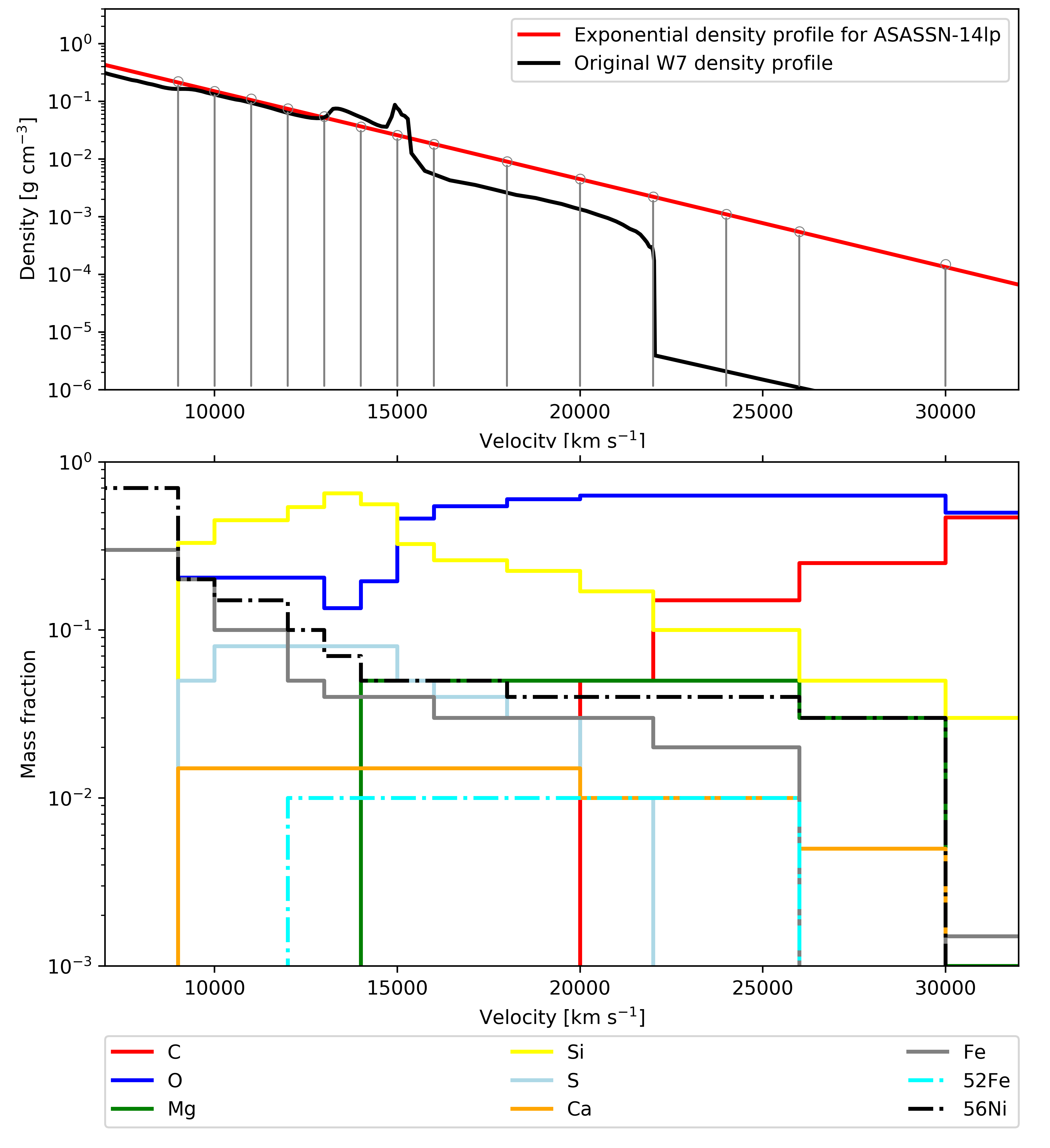}
    \caption{\textit{Top panel:} The best-fit pure exponential function for ASASSN-14lp (red) compared to the original W7 density structure at $t_\rmn{exp} = t_0 = 100$ s. The grey lines indicate the boundaries of the fitting layers for the corresponding chemical abundance models. \textit{Bottom panel:} The best-fit chemical abundance structure from the TARDIS fitting process of the ASASSN-14lp spectral time series. The profiles of the radioactive $^{52}$Fe and $^{56}$Ni show the mass fractions at $t_\rmn{exp}=100$ s.}
    \label{fig:abundance_sn14lp}
\end{figure*}

We find the date of the explosion as MJD 57\,997.05. This is in remarkably good agreement with the MJD 57\,996.93 estimated by \cite{Shappee16} is based on the fit of early-time expansion velocity evolution. In the case of the density profile, the adopted $\rho_0 = 5.0$ g cm$^{-3}$ and $v_0 = 2\,800$ km s$^{-1}$ characterize the exponential function very similar to the W7 density profile \citep{Nomoto84} below 15\,000 km s$^{-1}$, but contains significantly more mass in the outer regions, as can be seen in Fig. \ref{fig:abundance_sn14lp}. The figure also illustrates the chemical-abundance layers of the model.



The constrained best-fit abundance profiles of the elements listed in Sec. \ref{sec:fitting} can be also seen in Fig. \ref{fig:abundance_sn14lp}. Since the synthetic spectra are relatively insensitive to the abundances above 30\,000 km s$^{-1}$ due to the low densities, we keep the mass fractions fixed in this outermost region. We assume that the C/O dominated layer representing the unburnt material of the WD is also located above 30\,000 km s$^{-1}$ ($X(\rmn{O})=0.50$, $X(\rmn{C})=0.465$). Based on the case of SN 2013dy (see in \ref{appendix:sn2013dy}), $X(\rmn{Si})=0.03$ is used for this outermost layer to reproduce the early Si II HVF features \citep{Silverman15}. For the other chemical elements and isotopes considered in our analysis (see in Sec. \ref{sec:fitting}, solar metallicity was adopted. Note that our fitting strategy aims to constrain the chemical abundances only between 9\,000 ($v_\rmn{phot}$ at the latest epoch in our spectral sample) and 30\,000 km s$^{-1}$ (above which the chemical mass fractions are kept fixed).

Below 30\,000 km s$^{-1}$, the carbon abundance decreases constantly. Carbon cannot occur below 20\,000 km s$^{-1}$ in our models, otherwise the barely observable absorption of C II $\lambda$6580 would become too prominent. This kind of absence is predicted by several explosion scenarios, e.g. delayed detonation transition \citep{Ropke12, Seitenzahl13, Sim13}, gravitationally confined detonation \citep{Seitenzahl16} and double-detonation \citep{Fink10}, although, with various velocity limits. Our constraint is consistent with the hydrodynamic simulation of the spontaneous detonation of a 1.0 $M_\odot$ WD \citep{Shen18}, which predicts the allowed velocity region for carbon over 21\,000 km s$^{-1}$ \citep{Heringer19}. Meanwhile, oxygen remains the most abundant element much deeper into the ejecta in our best-fit model, with mass fraction $X(\rmn{O})=0.50-0.60$.

The main feature thought to explain the mid-UV flux suppression is the presence of IGEs at high velocities. By fitting the earliest spectra of ASASSN-14lp, we find that a significant mass fraction of $^{56}$Ni and its daughter isotopes are required even between 26\,000 and 30\,000 km s$^{-1}$ in our model ejecta. For the initial iron distribution, a slightly lower, 26\,000 km s$^{-1}$ velocity is required to fit Fe II and III features in the earliest spectra. The high IGE abundances at extremely high velocities deteriorate the fit of the nickel-cobalt feature at $\sim$3300 \r{A} and the Fe II lines between 4300 and 5000 \r{A}. However, these absorption features appeared also too strong in the synthetic spectra of SN 2013dy where moderate IGE abundances are inferred above 15\,000 km s$^{-1}$. Another effect, where the outer $^{56}$Ni abundance may be in conflict with the observations, is the impact on the early LC, which is investigated in Sec. \ref{sec:turtls}.

Both the mass fractions of stable Fe and $^{56}$Ni show a nearly constant trend with only a slight increase toward lower velocity, reaching 0.03 and 0.06 at 15\,000 km s$^{-1}$ respectively. Below that, IGE abundances start a more rapid increase, although IMEs dominate the ejecta. Within the scope of the analyzed ejecta region (> 9\,000 km s$^{-1}$), IGEs never become dominant in the model ejecta. However, based on the estimated total $M_\rmn{^{56}Ni}$ and the trends of the abundance profiles, we can assume that $X(^{56}\rmn{Ni})$ is the most abundant isotope below 9\,000 km s$^{-1}$. At the same time, the mass fraction of $^{52}$Fe used for controlling the contribution of manganese and chromium does not exceed 1\% in any region. Moreover, to avoid the detrimental effect of its decay product $^{52}$Cr on the last two spectral epoch, we eliminate it (i.e. set $X(^{52}\rmn{Fe})=0.0$) below 12\,000 km s$^{-1}$).

\begin{figure}
	\includegraphics[width=\columnwidth]{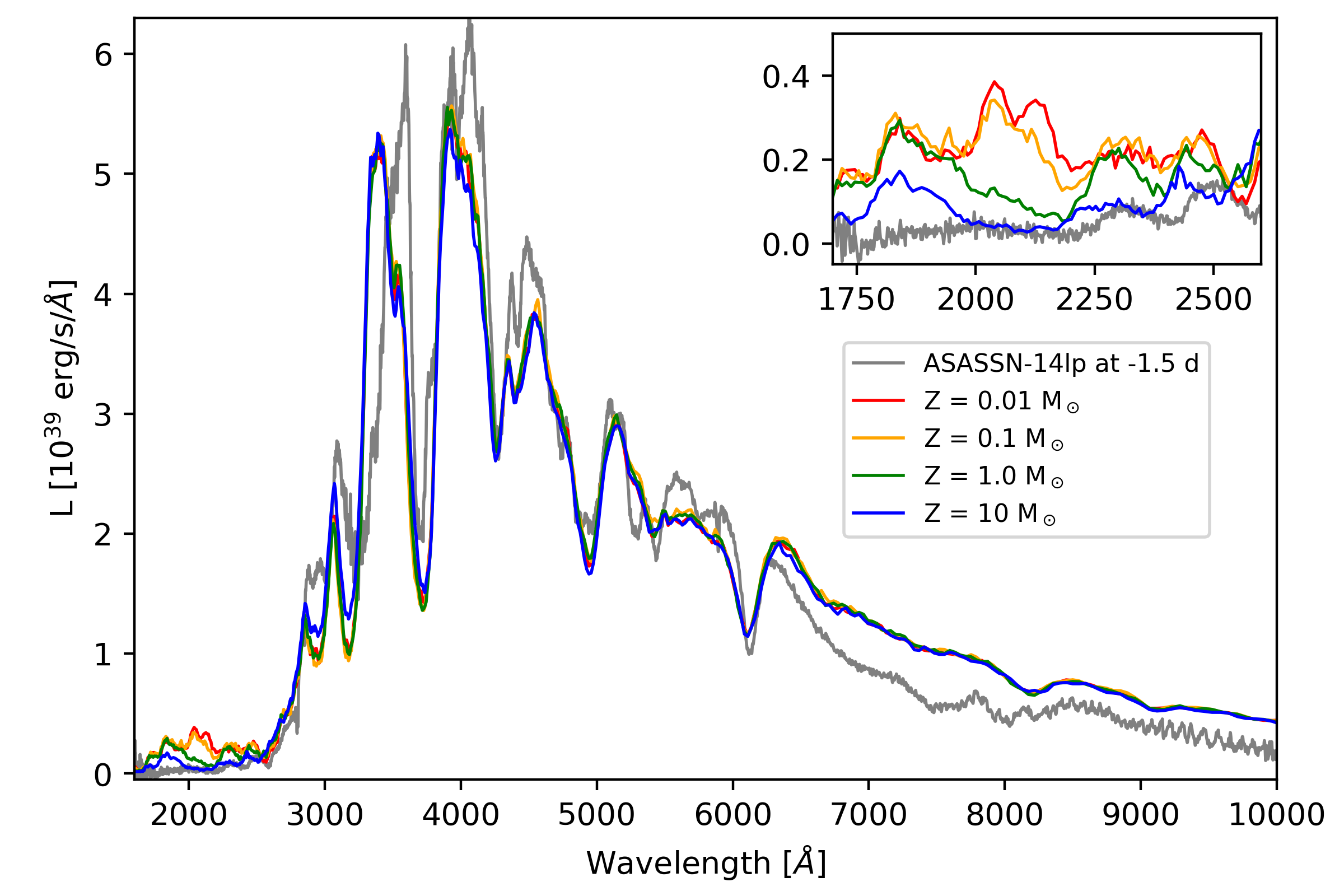}
    \caption{Model spectra with $X(^{56}\rmn{Ni}) = 0$ and different progenitor metallicities in the outermost fitting layer, i.e. between 15\,000 and 30\,000 km s$^{-1}$ in the model structure. The spectrum of ASASSN-14lp obtained at -1.5 days is also plotted (grey).}
    \label{fig:metallicity_test}
\end{figure}

\begin{figure}
	\includegraphics[width=\columnwidth]{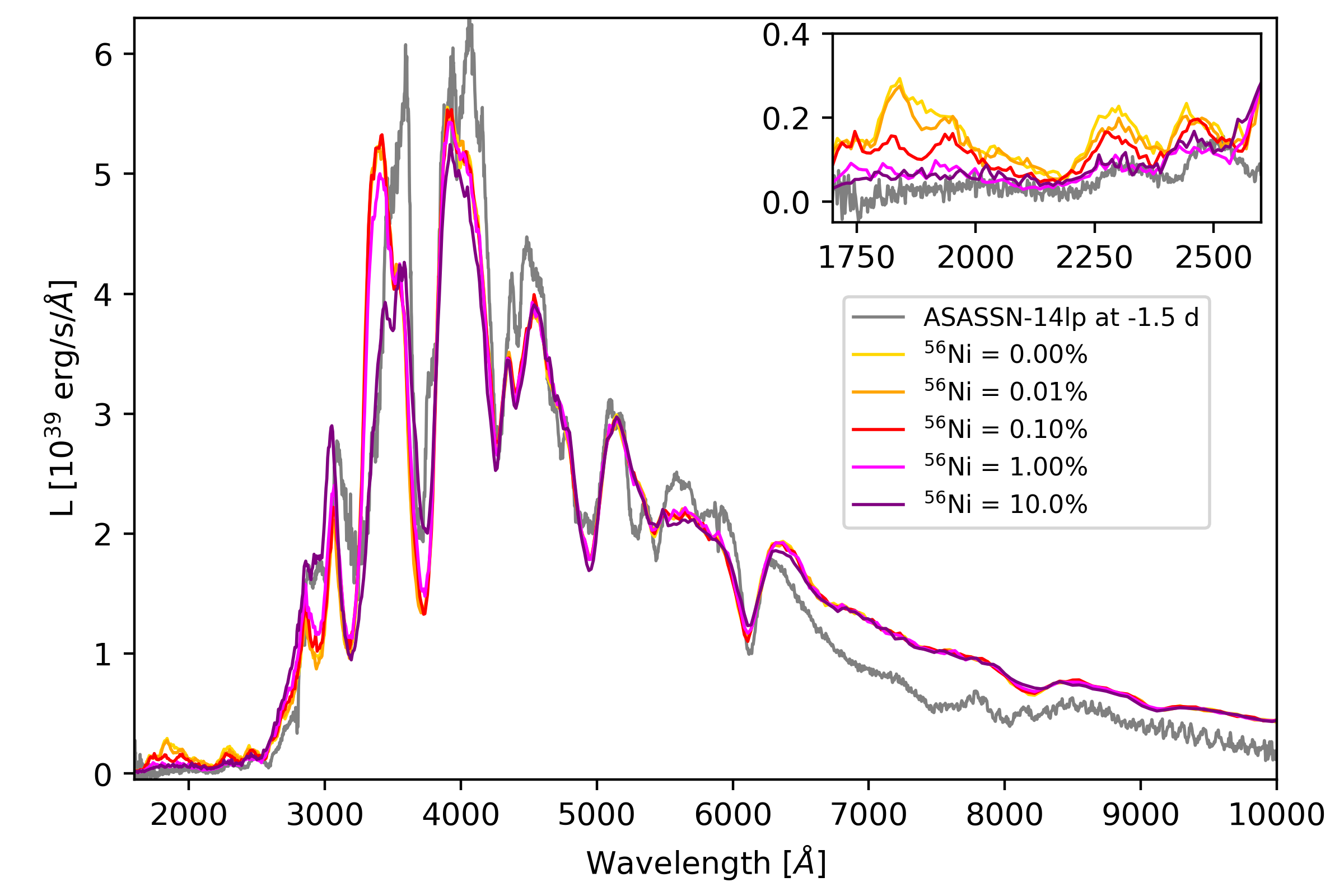}
    \caption{Model spectra with the same progenitor metallicity ($Z=Z_\odot$) and different mass fractions of $^{56}$Ni in the outermost fitting layer, i.e. between 15\,000 and 30\,000 km s$^{-1}$ in the model structure. The spectrum of ASASSN-14lp obtained at -1.5 days is also plotted (grey).}
    \label{fig:nickel_test}
\end{figure}

During the fitting process, we avoided rapid changes and peaks in the chemical abundance profiles and strived for constraining monotonic trends over the studied velocity range; however, in the case of IMEs like silicon and sulfur, the prominent lines clearly required a peak in the abundance profiles at intermediate velocities. The best-fit model structure contains $X(\rmn{Ca})=0.005 - 0.010$ and $X(\rmn{Si})=0.05 - 0.18$ in the region between 20\,000 and 30\,000 km s$^{-1}$, which reproduces the blue wings of the Ca II H\&K feature and the Si II $\lambda6355$ line, respectively. The strongly blueshifted absorptions of Ca II and Si II at early phases are typical of high-velocity features (HVFs), though the HVFs are not detached from the photospheric component in the case of ASASSN-14lp.
While the Ca abundance barely changes through the analyzed ejecta volume, the mass fraction of Si increases constantly toward the inner regions and reaches its peak with $X(\rmn{Si})=0.70$ between 13\,000 and 15\,000 km s$^{-1}$. Below that velocity, Si abundance seems to decrease slowly, which makes Si the most abundant chemical element in the region between 9\,000 and 15\,000 km s $^{-1}$. Considering that the synthetic spectra often show weaker Si II $\lambda$6355 absorption than the observed ASASSN-14lp spectra, the high Si abundance is probably not overestimated. Sulfur appears below 22\,000 km s$^{-1}$ in the best-fit model structure. From there on, S mimics the Si abundance profile with 6-8 times lower mass fraction.
Magnesium is also included in our best-fit model at the highest velocities, but the lack of unblended Mg II features makes the $X(\rmn{Mg})$ determination ambiguous. A magnesium mass fraction of 0.05 below $\sim$14\,000 km s$^{-1}$ does not further improve the fit, and we, thus, eliminate it from the inner regions.

\subsection{The impact of extended IGE distribution}
\label{sec:IGE_impact}

As highlighted in Sec. \ref{sec:introduction}, the distribution of IGEs strongly affects the spectra. The UV-suppression is particularly sensitive to the IGE abundances in the outer region of the ejecta. There are two main sources that may account for the amount of IGEs at early epochs, the unburnt material of the progenitor metallicity and the products of the explosive nucleosynthesis (mainly $^{56}$Ni). Note that the abundance tomography of ASASSN-14lp resulted in 3-4\% of $^{56}$Ni between 15\,000 and 30\,000 km s$^{-1}$, but our analysis did not fit different progenitor metallicities. In order to study the impact of IGE abundances at high velocities, we performed comprehensive tests for the near-maximum spectrum of ASASSN-14lp. For this, we adopt the best-fit TARDIS model and vary the chemical composition in the outer fitting region (i.e. between 15\,000 and 30\,000 km s$^{-1}$). Abundances in below and above this region remained the same as in our best-fit TARDIS model. 

At first, the mass fraction of $^{56}$Ni was set to 0 in the outer region, and we computed synthetic spectra varying the progenitor metallicity by three orders of magnitude (see in Fig. \ref{fig:metallicity_test}). We found that the observed UV suppression can be reproduced only with at least ten times the solar metallicity. Such an extreme abundance of heavier elements has never been found in WD spectra. In conclusion, we discard the idea to fit this spectral region by scaling the progenitor's metallicity.


A similar approach was chosen to test the impact of radioactive nickel in the outermost fitting region, varying the $^{56}$Ni mass fraction by orders of magnitude (see Fig. \ref{fig:nickel_test}). As expected \citep{Brown15}, the largest impact was on the suppression of the UV flux. The model structures with $X(^{56}\rmn{Ni}) = 0.001$ or higher in the velocity region between 15\,000 and 30\,000 km s$^{-1}$ reproduced the observed flux level below 2800 \r{A}. However, with increasing $^{56}$Ni mass fraction, the models systematically deviate in the range of the strong absorption profile between 3100 and 3500 \r{A}, and the corresponding pseudo-emission between 3500 and 3900 \r{A}. In conclusion, the extended $^{56}$Ni distribution seems a feasible solution to reproduce the UV part of the spectra.

\subsection{Constraints on the allowed helium shell in a double detonation scenario}
\label{sec:he_shell}

While the chemical abundance structures presented here show a good agreement in general with former modeling approaches \citep[e.g.][]{Stehle05, Mazzali08}, the results related to the outermost regions ($v > 20\,000$ km s$^{-1}$) might be questionable. Note that the photosphere is formed around 16\,000 km s$^{-1}$ at the earliest spectral epoch, thus, the layers above that are not tested directly by fitting the whole spectral range. The abundance values of the outermost regions are constrained based on the fitting of the mid-UV wavelengths and the blue wings of strong absorptions features. IMEs, like Si and Ca, may appear with significant mass fractions between 20,000 and 30\,000 km s$^{-1}$ in the ejecta, which is supported by the formation of the high velocity features \citep{Silverman15}. At the same time, explosion models predict typically no IGEs at extremely high velocities \citep[see e.g.][]{Hillebrandt13}. As we highlight in the Sec. \ref{sec:introduction}, the only promising explosion scenario with significant $^{56}$Ni mass in the outer layers is the family of the double detonation models, as reported by several hydrodynamic simulations \citep[][]{Kromer10,Fink10,Woosley11,Sim12}. A few examples of SNe Ia (or peculiar Type I SNe) have been linked so far to the DD scenario, e.g. SNe 2018byg and 2019vyq, in case of which a limit was set on the helium shell based on the estimated $^{56}$Ni yields.

Here, we briefly compare our results on ASASSN-14lp to the estimated radioactive nickel yields of the hydrodynamic models presented in the literature. Such comparison cannot be detailed because i) we have no information from below the photosphere of the last epoch, i.e. about the inner regions at $v < 8\,500$ km s$^{-1}$, and total ejecta and $^{56}$Ni masses cannot be constrained; ii) the distribution of the outermost $^{56}$Ni is poorly constrained (see the previous paragraph). Thus, the comparison is limited only to the mass of the pre-explosion He-shell $M_\rmn{shell}$ and the $^{56}$Ni produced in the shell-detonation. The fusion products of the He-shell detonation are assumed to be located over 15\,000 km s$^{-1}$ in the model ejecta of ASASSN-14lp, right above the region dominated by IMEs. This part of ejecta has the total mass of 0.066 M$_\odot$ and consists of 0.003 M$_\odot$ of $^{56}$Ni.

\cite{Sim12} assumed two WD progenitor with M$_\rmn{He} = 0.21$ M$_\odot$ to simulate convergence-shock and edge-lit double detonation scenarios for both setups in 2D. Due to the low-mass C/O cores and the resulted kinetic energies, the IGE fusion products of the He-shell burning appear between 10\,000 and 20\,000 km s$^{-1}$. From the small sample of models, ASASSN-14lp favours the less massive setup with the converging-shock scenario resulting in a M$_\rmn{^{56}Ni} = 0.0028$ M$_\odot$ in the outer layers.

In the hydrodynamic simulations of \cite{Fink10}, six different initial C/O WD (between 0.81 and 1.385 M$_\odot$) setup were assumed with minimum masses of accreted He-shells that could lead to a detonation (M$_\rmn{He} = 0.126 -- 0.0035$ M$_\odot$). As a common point, the detonations of the He-shell produce an IGE-peak in the abundance structure, which rapidly descents toward the outer regions. This feature contradicts with our picture of ASASSN-14lp, where 1 – 4 \% of $^{56}$Ni abundance above 20000 km s$^{-1}$.  The inferred $^{56}$Ni yield shares the most similarities with Model 3 and 4, predicting $M_\rmn{^{56}Ni} = 0.0017$ and 0.0044 $M_\odot$ as the fusion product of M$_\rmn{He} = 0.055$ and $0.039$ M$_\odot$, respectively. However, only models with small helium shell mass ($M_\rmn{He} < 0.04 M_\odot$) show high values of $X(\rmn{IGE})$ over 20\,000 km s$^{-1}$, which vary depending on the angle of view and the strength of the detonation.

\cite{Polin19} extended the hydrodynamic simulations to a larger grid of He-shell and WD masses producing a several order of magnitude range of $^{56}$Ni produced in the shell. Five setups resulted in a similar M$_\rmn{^{56}Ni} = 1.5 - 6.4 \times 10^{-3}$ M$_\odot$ as the fusion product of a 0.02 - 0.08 M$_\odot$ He-shell. This comparison clearly indicates that the inferred $^{56}$Ni yields of ASASSN-14lp are in good match with realistic simulations of the double-detonation scenario. The detailed comparison with the predictions of the double detonation scenario is beyond the scope of this current paper and will be the subject of a future study.

\subsection{Early light curve analysis with TURTLS}
\label{sec:turtls}

The impact of IGE at extremely high velocities to the light curve evolution has to be also investigated. Based on our abundance tomography analysis, we find that IGE in the outermost ejecta can reproduce the flux suppression observed at UV wavelengths. The presence of such material, however, could significantly impact the light curve evolution. As TARDIS only provides snapshot spectra and requires the SN luminosity as input, it is not suited for investigating the impact of the IGEs on the light curves. For this, we use the one-dimensional, Monte Carlo radiative transfer code TURTLS \citep{Magee18}. TURTLS and TARDIS operate under the same principles of Monte Carlo radiative transfer adopted from \cite{Lucy05} and TARDIS is used to calculate the opacity of each grid cell in TURTLS. One of the key differences between TURTLS and TARDIS, is that in TURTLS packets are injected following the decay of $^{56}$Ni and there is no assumed photosphere. Therefore TURTLS allows for the calculation of light curves; however, it is ill-suited for rapid spectral modelling, in contrast to TARDIS, as packets must propagate throughout the entire ejecta.

We use the abundances derived from our abundance tomography analysis of ASASSN-14lp for the TURTLS simulations. We directly import the density profile and the abundance structure above 9\,000 km s$^{-1}$ from the best-fit TARDIS model. However, TURTLS requires information over the whole ejecta volume, which makes further assumptions necessary. The exponential function of the density profile is simply extrapolated towards the core of the ejecta, which is probably a realistic simplification considering the results of the hydrodynamic simulations of double-detonations \citep{Fink10,Sim12} or delayed-detonations \citep{Seitenzahl13}. For the abundance structure, we use only Fe and $^{56}$Ni abundances with constant mass fractions in the ejecta volume between 0 and 9,000 kms$^{-1}$. This choice is supported by the increasing abundance of these two species in deeper layers in our abundance tomography, the IGE dominance at low velocities in the related hydrodynamic simulations, and the total radioactive nickel mass ($M_\rmn{^{56}Ni}$) estimated through the bolometric LC. However, we have only indirect estimates of $M_\rmn{^{56}Ni}$ (strongly dependent on the assumed distance), thus, we adopt various levels for the inner mass fraction of $^{56}$Ni (for summary of the used $X(^{56}\rmn{Ni})$ in this section, look at Tab. \ref{tab:turtls_abundances}). The assumed constant mass fraction of $^{56}$Ni is set as $X_\rmn{inner}(^{56}\rmn{Ni})=0.50$, 0.65 and 0.80, which give total masses of $M_\rmn{tot}(^{56}\rmn{Ni})=0.49$, 0.62 and 0.74 M$_\odot$ respectively. 

\begin{figure}
	\includegraphics[width=\columnwidth]{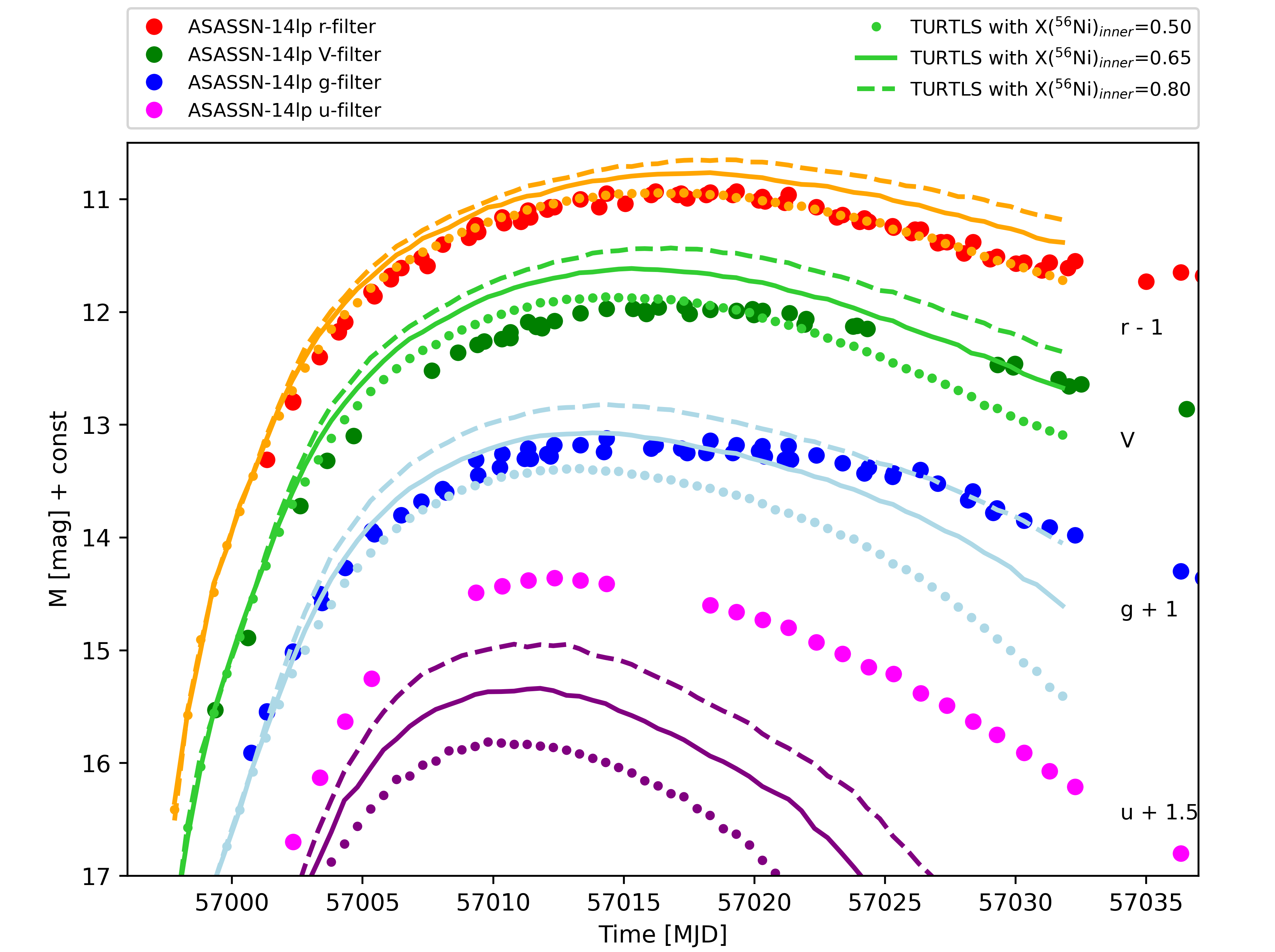}
    \caption{Comparison of the early light curves of ASASSN-14lp and the synthetic light curves in V (green), u (purple), g (blue) and r band (red) estimated with TURTLS based on the best-fit TARDIS model. The different symbols of the TURTLS light curves refer to different assumed mass fractions of $^{56}$Ni in the inner ejecta.}
    \label{fig:turtls}
\end{figure}

The resulting LCs are shown in Fig. \ref{fig:turtls}. The synthetic LCs are terminated at 35 days after the explosion because the adopted assumptions of local thermodynamic equilibrium are no longer reliable after that time. Note that our abundance tomography analysis also stops at $t_\rmn{exp}=26.4$ days (i.e. MJD 57\,023.4). The $^{56}$Ni of the inner regions has an impact on the LC before and around maximum light. Neither of the above mentioned $^{56}$Ni distributions is able to fully explain the observed LCs. However, the model with the lowest $^{56}$Ni mass shows an excellent match with the LC obtained in the $r$-filter and has only a slight discrepancy compared to the V-band LC, but strongly deviates from the $g$-band LC after the peak. The TURTLS model with an inner abundance of $X_\rmn{inner}(^{56}\rmn{Ni})=0.80$ fails to meet with the observed LCs, especially in the $V$-filter, where the discrepancies reach $\sim$1 mag. The intermediate model shows a moderately good agreement with all the three studied LCs during the first $\sim$30 days. Thus, we choose $X_\rmn{inner}(^{56}\rmn{Ni})=0.65$ as the inner abundance for further investigations below.

However, all the \textit{u}-band TURTLS LCs show a large deviation from the observed ones, indicating that the $^{56}$Ni masses assumed for this test are not sufficient to power the short-wavelength LC. Indeed, neither of the assumed mass fraction of $^{56}$Ni for the inner regions gives the $M_\rmn{tot}$($^{56}$Ni) = 0.806 estimated by \cite{Shappee16}. As the synthetic \textit{u}-band LCs are highly sensitive to the mass of radioactive material enclosed at low velocities, which were not sampled by our spectral analysis, the near-UV LCs are not suitable for testing the inferred abundance structure.

\begin{figure}
	\includegraphics[width=\columnwidth]{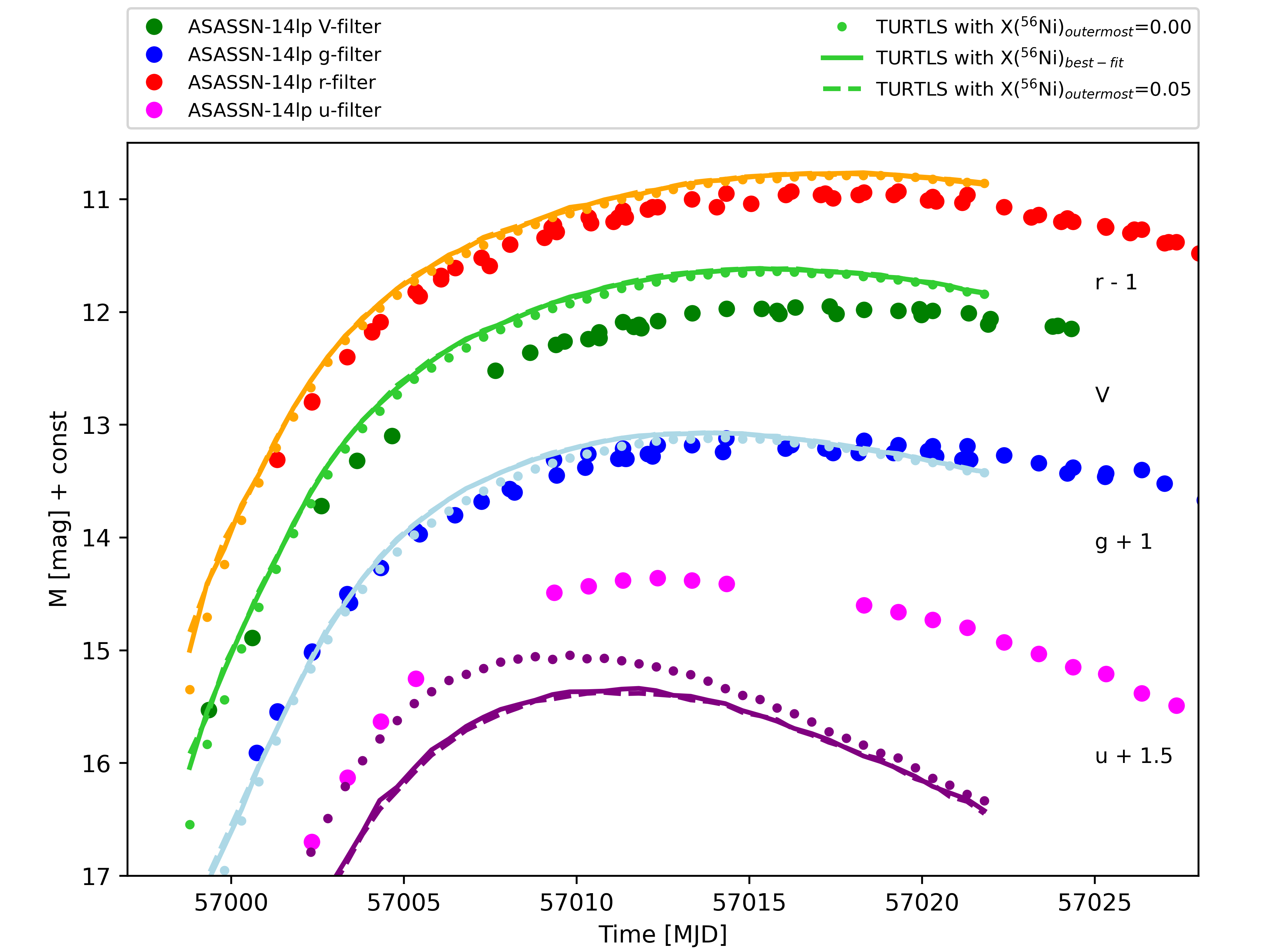}
    \caption{Comparison of the early light curves of ASASSN-14lp and the synthetic light curves in V (green), u (purple), g (blue) and r band (red) estimated with TURTLS based on the best-fit TARDIS model. The different symbols of the TURTLS light curves refer to different mass fractions of $^{56}$Ni in the top of the ejecta ($>18\,000$ km s$^{-1}$).}
    \label{fig:turtls1}
\end{figure}

\begin{figure}
	\includegraphics[width=\columnwidth]{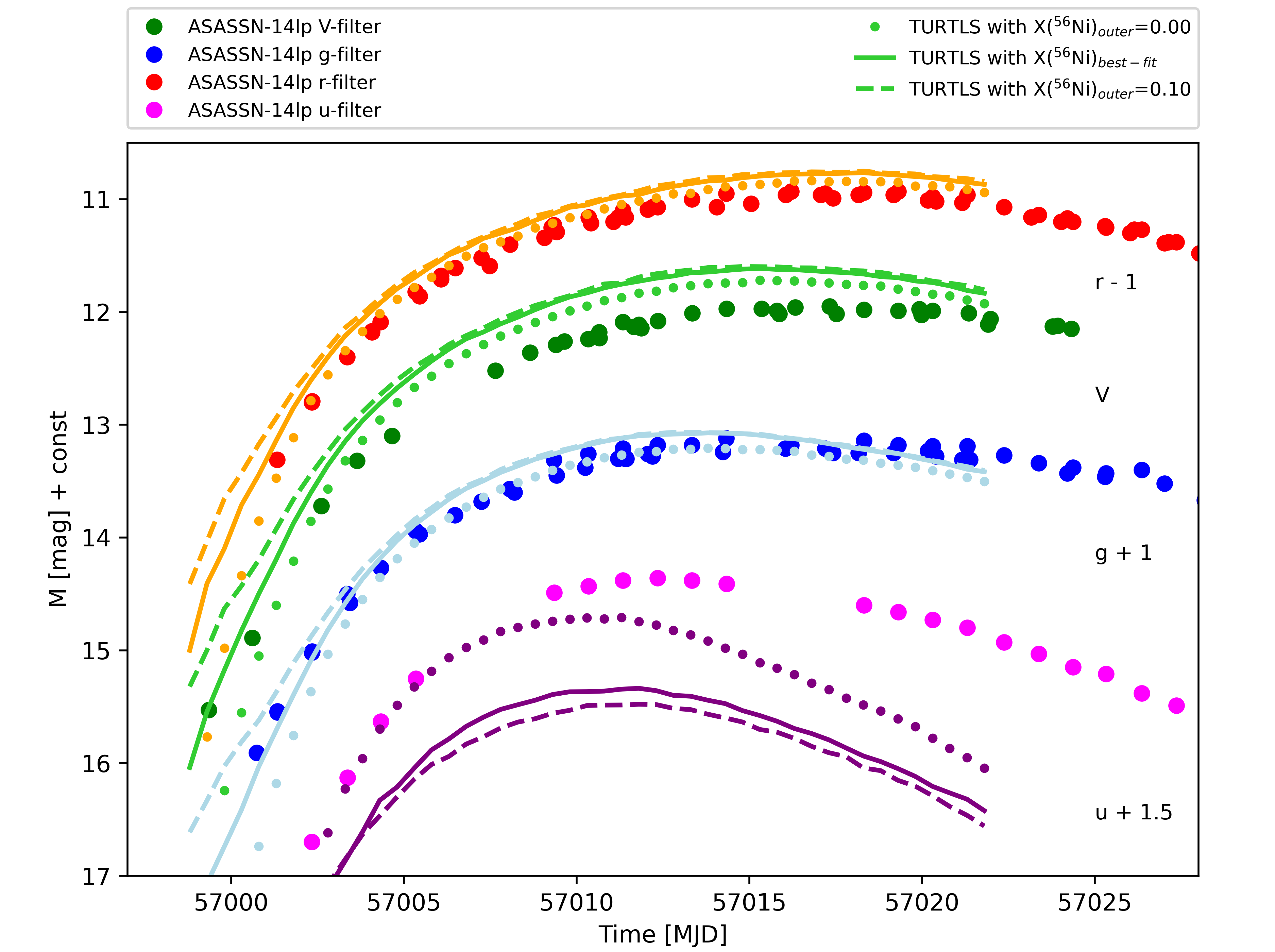}
    \caption{Comparison of the early light curves of ASASSN-14lp and the synthetic light curves in V (green), u (purple), g (blue) and r band (red) estimated with TURTLS based on the best-fit TARDIS model. The different symbols of the TURTLS light curves refer to different mass fractions of $^{56}$Ni in the outer regions of the ejecta ($>12\,000$ km s$^{-1}$).}
    \label{fig:turtls2}
\end{figure}

\begin{table*}
	\centering
	\caption{Summary of the variation of X($^{56}$Ni) in the comparison tests performed with TURTLS in Sec. \ref{sec:turtls}. The 'best-fit' label refers to the corresponding mass fraction of $^{56}$Ni from the best-fit TARDIS model, assuming X($^{56}$Ni) = 0.04 - 0.10, and 0.00 - 0.04 in the outer and top regions, respectively.}
    \label{tab:turtles}
    \begin{tabular}{|c|ccc|ccc|ccc|}
		\hline
		& \multicolumn{3}{c}{inner region} & \multicolumn{3}{c}{outer region} & \multicolumn{3}{c}{top region}\\
		& \multicolumn{3}{c}{v < 9\,000 km s$^{-1}$} & \multicolumn{3}{c}{12\,000 < v < 18\,000 km s$^{-1}$} & \multicolumn{3}{c}{v > 18\,000 km s$^{-1}$}\\
		\hline
		Line style in figures & dotted & solid & dashed & dotted & solid & dashed & dotted & solid & dashed \\
		\hline
        Comparison in Fig. \ref{fig:turtls} & 0.50 & 0.65 & 0.80 &      \multicolumn{3}{c}{best-fit (0.04-0.10)}     &  \multicolumn{3}{c}{best-fit (0.00-0.04)}       \\
        Comparison in Fig. \ref{fig:turtls1} & 0.65 & 0.65 & 0.65 & & best-fit & & 0.00 & best-fit & 0.05\\
        Comparison in Fig. \ref{fig:turtls2} & 0.65 & 0.65 & 0.65 & 0.00 & best-fit & 0.10 & 0.00 & best-fit & 0.10\\
        \hline
	\end{tabular}
	\label{tab:turtls_abundances}
\end{table*}

We test the impact of the $^{56}$Ni distribution from the best-fit model in two steps. First, we modify the $^{56}$Ni abundance above 18\,000 km s$^{-1}$ (the \textit{top} of the the ejecta, hereafter) and set constant mass fractions of $X_\rmn{top}(^{56}\rmn{Ni})=0.00$ and 0.05 (note that the abundance varies between 0.03 and 0.05 in the best-fit TARDIS model). The TURTLS synthetic LCs compared to those based on the best-fit model and the observed photometry can be found in Fig. \ref{fig:turtls1}. The synthetic LCs of the best-fit and $X_\rmn{top}(^{56}\rmn{Ni})=0.05$ models are indistinguishable, while the LCs for $X_\rmn{top}(^{56}\rmn{Ni})=0.00$ differ by less than 0.1 mag. We can assume that $^{56}$Ni abundance at extremely high velocities does not have a significant impact on the LC if the mass fraction stays below 5\%. In conclusion, the chemical abundances of the best-fit TARDIS model, which play a major role in the formation of the mid-UV suppression, are not in conflict with the observed photometry.

As a final comparison (in Fig. \ref{fig:turtls2}), we also calculate TURTLS synthetic LCs for models with modified abundance profiles between 12\,000 and 18\,000 km s$^{-1}$ (in the  \textit{outer} region, hereafter), having constant mass fraction of $X_\rmn{outer}(^{56}\rmn{Ni})=0.00$ and 0.10. The same abundances are adopted for $X_\rmn{top}(^{56}\rmn{Ni})$ (see in Tab. \ref{tab:turtls_abundances}). The resulting deviations from the LCs of the best-fit model are usually small ($\sim$0.2 mag) except before MJD 57\,003, when the contribution of the outer regions to the LC is higher. At these early epochs, the observed LCs show a marginally better agreement with the original best-fit abundance profile. Note that the lack of photometry before MJD 57\,000 ($t_\rmn{exp}=3$) days hinders such an analysis, and the formation of the LCs after MJD 57\,006 is clearly dominated by the inner regions, which is unconstrained from our abundance tomography study.

\subsection{The impact of the photospheric SED}
\label{sec:photosphere}

In this section, we test whether the application of such an extreme IGE distribution can be avoided with reasonable modifications of the physical assumptions. One of the main simplifications of TARDIS is the assumption of a sharp photosphere at the inner boundary of the modelling volume that radiates like a blackbody \citep{Kerzendorf14}. 

There are indications in the literature \citep[e.g.][]{Pinto00} that the assumption of such a photosphere might not be robust in the case of thermonuclear SNe and the radiation field emerging from the inner layers may deviate from a blackbody ($B(\lambda,T_\rmn{phot})$). The main source of this effect is the enhanced blanketing of cobalt and nickel ions \citep{Dessart14}, which represent the majority of the total mass below 10\,000 km s$^{-1}$, thus, one can expect a strong flux suppression at the characteristic wavelengths of these ions. As a more realistic alternative, the radiation transport from the inner layers should be modelled, which could be adopted as the spectral energy distribution at the inner boundary (hereafter $I_\rmn{phot}(\lambda)$) of a TARDIS simulation.

To illustrate a possible $I_\rmn{phot}(\lambda)$ emerging from the innermost layers, we calculated the emission at different depths in the model ejecta of ASASSN-14lp with ARTIS \citep{Kromer09}. Although ARTIS is based on the same principles of radiative transfer as TARDIS, it offers a more complex approach to synthesize model spectra.
ARTIS calculates the propagation of photon packages from the moment of radioactive decay through the whole ejecta, instead of assuming an inner boundary at which a radiation field is injected. Modelling the diffusion of radiation at high optical depth comes with a high computational cost, thus, ARTIS is not used for spectral fitting purposes.

\begin{figure}
	\includegraphics[width=\columnwidth]{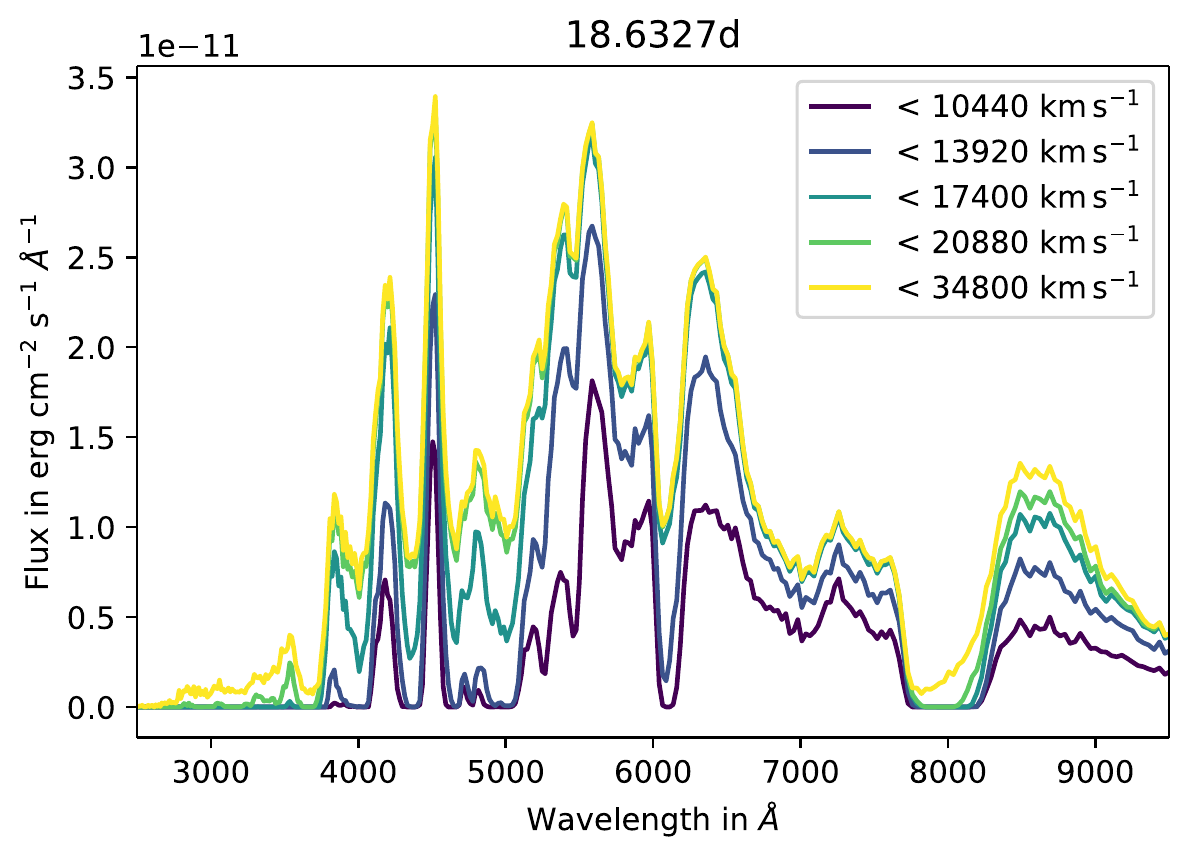}
    \caption{Synthetic spectra calculated with ARTIS assuming the density profile and abundance structure of the best-fit TARDIS model for the epoch +0.3 day. The different colours represent the emerging radiation from spheres at different velocities.}
    \label{fig:artis}
\end{figure}

For the ARTIS simulation, we use the same ejecta structure as in the previous section. We assume $X(^{56}\rmn{Ni})=0.65$ and $X(\rmn{Fe})=0.35$ for the layers below 9\,000 km s$^{-1}$, where the abundance tomography provided no information on the chemical composition. Above 9\,000 km s$^{-1}$, we adopt the best-fit abundances from the TARDIS fitting. For the epoch at $t_\rmn{exp}=18.6$ days, the flux emerging from different layers in the ARTIS simulation can be seen in Fig. \ref{fig:artis}. The emerging radiation at $v=10\,440$ k s$^{-1}$ (close to the assumed photosphere at $t_\rmn{exp} = 18.6$ days) deviates significantly from a $B(\lambda,T_\rmn{phot})$ assumed in our TARDIS models. The discrepancy is particularly strong between 4500 and 5500 \r{A} and at the UV wavelengths, where the flux suppression can be observed without the contribution of \textit{any} IGEs in the outer regions. The latter could have particularly high importance for our analysis: could an alternate photospheric SED reproduce the UV suppression without an extended source of $^{56}$Ni?

In order to test the impact of $I_\rmn{phot}(\lambda)$, we compare two cases: (i) $I_\rmn{phot}(\lambda) = B(\lambda,T_\rmn{phot})$, which is the standard choice for TARDIS, and (ii) $I_\rmn{phot}(\lambda) = \Theta(\lambda - \lambda_\rmn{cut}) \times B(\lambda,T_\rmn{phot})$, where $\Theta$ is the Heaviside jump function and we assume $\lambda_\rmn{cut}=3000$ \r{A}. Otherwise, the modelling of the propagation and interaction of the photon packages are left untouched in TARDIS. Note that the truncation of incoming flux in the UV regime at the photosphere does not necessarily result in a lack of observed flux at these wavelengths, because the {\tt macroatom} method for line interaction in TARDIS simulations \citep{Lucy02,Lucy03} could redistribute the photon packages even to shorter wavelengths.

\begin{figure}
	\includegraphics[width=\columnwidth]{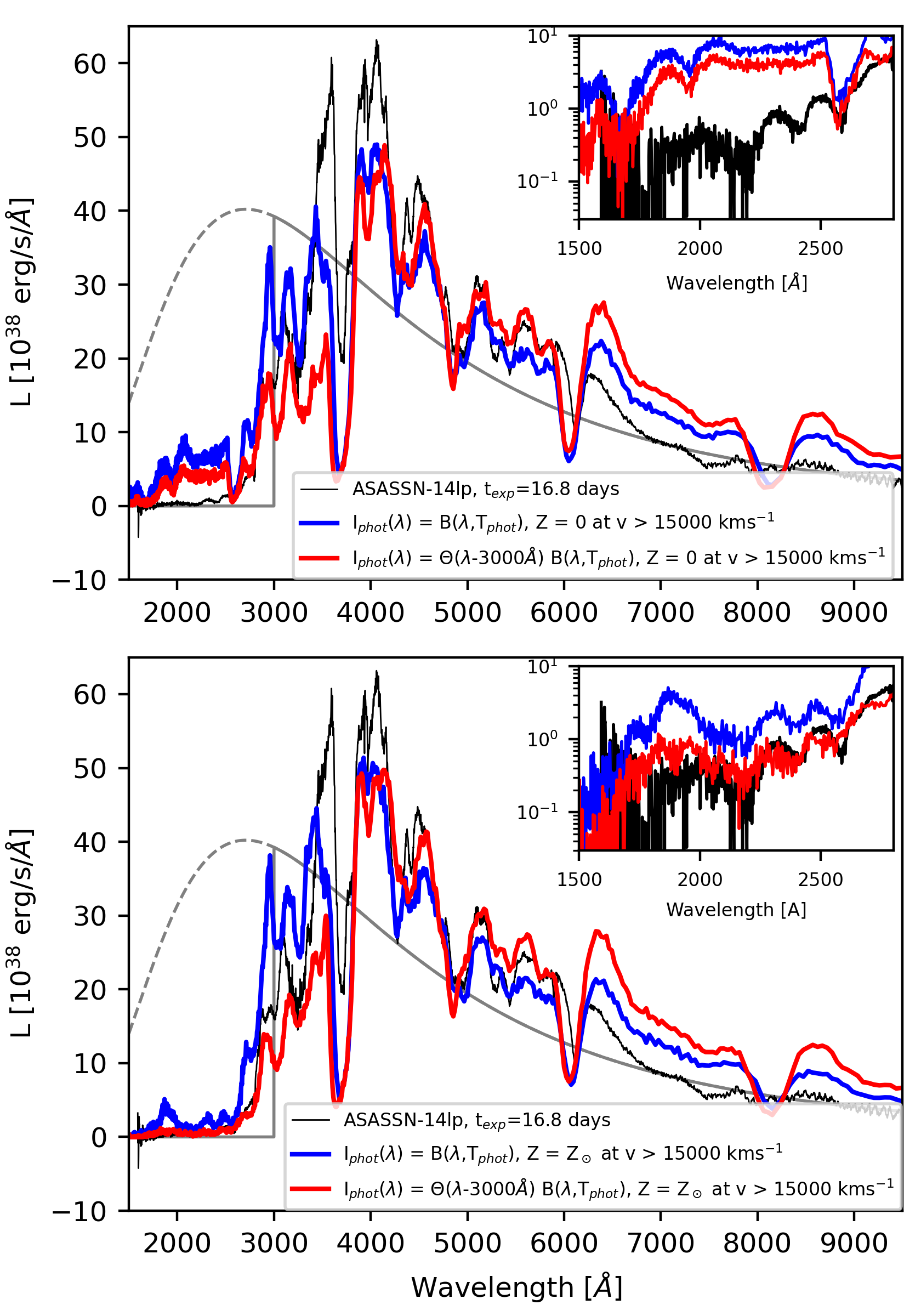}
    \caption{Comparison between adopting a normal Planck function and one truncated below 3\,000 \r{A} as radiation field at the photosphere in the TARDIS model of ASASSN-14lp. The chemical composition of the best-fit model is changed above 15\,000 km s$^{-1}$, using mass fractions of X(IGE) = 0.0 (top panel) and X(IGE) = Z$_\odot$ (bottom panel) in both cases. The corresponding spectrum of ASASSN-14lp at the epoch -1.5 days with respect to $B$-band maximum (black) is also shown. The grey lines represent the $I_\rmn{phot}(\lambda)$ at the inner boundary, for the normal (dashed) and the  modified (solid) Planck function.}
    \label{fig:truncated}
\end{figure}

The impact of the modified $I_\rmn{phot}(\lambda)$ is demonstrated on the ASASSN-14lp spectrum taken at $t_\rmn{exp}=16.8$ days. We examine two configurations: one without any Fe and Ni between 15\,000 and 30\,000 km s$^{-1}$, and one with solar mass fractions for the two chemical elements. We focus only on the UV part of the spectrum, thus, the rest of the abundance structure is directly adopted from the best-fit TARDIS model. The results of the test can be seen in Fig. \ref{fig:truncated}. Since the TARDIS {\tt macroatom} scheme allows for a fully general treatment of line formation including (reverse-)fluorescence processes, the output spectrum is not truncated below 3000 \r{A}. Moreover, the observed UV suppression does not appear when zero metallicity assumed. At the same time, assuming solar mass fractions of the IGEs, the truncated $I_\rmn{phot}(\lambda)$ clearly reproduces the low flux level in the UV with similar efficiency as the extended $^{56}$Ni abundance in the best-fit TARDIS model.

Considering our limited knowledge about the radiation field that emerges from the inner layers, the modification of the inner boundary $I_\rmn{phot}(\lambda)$ for the abundance tomography would be an arbitrary choice; thus, investigating its full impact on the abundance tomography is out of the scope of this study. Furthermore, the characterization of the $I_\rmn{phot}(\lambda)$ would introduce too many new parameters in the fitting with high degeneracy.
In a conclusion, we cannot claim that \textit{only} the truncated $B(\lambda,T_\rmn{phot})$ is responsible for the observed UV suppression, but as we showed, such $I_\rmn{phot}(\lambda)$ could reproduce the observed UV features with a more realistic abundance profile of the outer ejecta.

\section{Conclusions}
\label{sec:conclusions}

We performed spectroscopic analysis on  ASASSN-14lp, a normal Type Ia SN showing strong flux suppression in the near UV-regime. Our study aimed to investigate whether this feature can characterize the outer ejecta. As a conclusion, we presented two possible origins.

The UV suppression previously thought to be the effect of increased metallicity or a significant amount of radioactive $^{56}$Ni in the outer regions of the ejecta \citep{Brown15}. We constrained the chemical abundance structure of ASASSN14lp via the technique called abundance tomography \citep{Stehle05}. The spectral time series obtained before and around maximum light was fit with synthetic spectra calculated from a self-consistent model ejecta. The structure of the model ejecta was defined with the density profile and the chemical abundances as functions of velocity. The synthetic spectra were computed with the one-dimensional Monte Carlo radiative transfer code TARDIS \citep{Kerzendorf14}. Our fitting method mainly followed the strategy described in \cite{Barna18}. 


The best-fit synthetic spectra reproduce the positions, strengths, and time evolution of the observed spectral features well. Apart from the strong absorption features, the fitting process focused on the mid-UV regime. A preliminary test showed that only extreme metallicities of about $Z = 10 Z_\odot$ resulted in the observed level of flux, which is an unreasonably high value considering the nearly solar metallicities of the progenitor objects. A significant $^{56}$Ni mass fraction of $X(^{56}\rmn{Ni}) = 0.01-0.05$, however, is able to reproduce the flux deficit, if it extends all the way to the outermost layers at 25\,000 - 30\,000 km s$^{-1}$. Relatively high $^{56}$Ni abundance at large velocities is only expected in specific explosion scenarios, like the double-detonation \citep{Kromer10,Woosley11} or delayed-detonation \citep{Sim13}, however, radiative transfer modelling shows severe deviations from observed SNe. Thus, the existence of the inferred $^{56}$Ni distribution is questionable. 

Since the extended $^{56}$Ni abundance may have a strong impact on the early LCs, we used the radiative transfer code TURTLS \citep{Magee18} to test the best-fit model structure of ASASSN-14lp against its observed magnitudes in different filters. This simulation required information about the ejecta even below 9\,000 km s$^{-1}$, which was not provided by our abundance tomography analysis. Thus, we assumed different constant mass fractions of $^{56}$Ni in these inner regions. We found that the different inner abundances have a dominant impact on the LCs after $t_\rmn{exp}=5$ days, preventing the clear investigation of the best-fit $^{56}$Ni abundance in the outer layers by this method. None of the estimated synthetic LCs was able to reproduce all of the observed LCs. However, the model with an inner $^{56}$Ni abundance of $X(^{56}\rmn{Ni})=0.65$ shows a relatively good agreement with both $g$-, $V$- and $r$-band LCs with maximum discrepancies of $\sim$0.5 mag. By modifying the abundances above 18\,000 km s$^{-1}$ we found that these top layers do not affect significantly the produced LCs if the mass fraction of $^{56}$Ni is low ($X_\rmn{top}(^{56}\rmn{Ni})<0.05$). A similar test was performed for the outer layers over 12\,000 kms$^{-1}$, which leads to the conclusion that the $^{56}$Ni distribution of our best-fit model is not in tension with the observables.

Besides $^{56}$Ni, the abundance profiles of additional seven chemical elements, namely C, O, Mg, Si, S, Ca, Fe, and the radioactive isotopes $^{52}$Fe were constrained, also considering the decay products of the radioactive isotopes. Some of the abundance profiles have high uncertainty because of the low mass fractions (e.g. Ca) or the lack of any unambiguous absorption features (Mg, O). The appearance of carbon is limited below 20\,000 km s$^{-1}$ in the model ejecta. The mass fractions of silicon and sulfur change in a similar way with the only difference that silicon appears even at the highest velocities above 25\,000 km s$^{-1}$. The mass fractions of these IMEs peak around 12\,000 - 14\,000 km s$^{-1}$, where silicon is the most abundant element over a narrow velocity range. The stable iron abundance follows the changes of $X(^{56}\rmn{Ni})$ with an increasing trend toward the inner regions.

As an alternative explanation for the UV suppression, we also investigated whether the lack of flux in the UV regime could emerge from deeper layers as indicated in ARTIS \citep{Kromer09} simulations. In order to test this possibility, we adopted a modified $I_\rmn{phot}(\lambda)$ at the inner boundary of the TARDIS simulation, which does not contain any photon package with a wavelength lower than 3000 \r{A}. The lack of UV packages at the inner boundary of the simulation does not automatically result in a fully truncated spectrum, because of the reverse fluorescence process. As shown with a model ejecta having $X(\rmn{IGE}) = 0.0$ above 15\,000 km s$^{-1}$, the UV-suppression is not reproduced with zero metallicity in the outer regions. However, this part of the spectrum remains very sensitive to the amount of IGEs even in the case of the modified $I_\rmn{phot}(\lambda)$. By adopting only solar metallicity (i.e. $X(\rmn{IGE}) = Z_\odot$) in the layers above 15\,000 km s$^{-1}$, the synthetic spectrum showed a good match with the observed UV suppression. Thus, using IGE abundances at extremely high velocities in our ejecta model can be avoided by relaxing the assumption of a blackbody radiation field at the inner boundary.

Whether or not this assumption on the $I_\rmn{phot}(\lambda)$ is more realistic than the excessive distribution of radioactive Ni constrained in our abundance tomography analysis, is out of the scope of this study and requires further investigation.

\section*{Acknowledgements}

This work is part of the project Transient Astrophysical Objects GINOP-2-3-2-15-2016-00033 of the National Research, Development and Innovation Office (NK-FIH), Hungary, funded by the European Union. BB received support by the Hungarian NKFIH/OTKA FK-134432 grant, by the Czech Science Foundation grant 19-215480S, and by the institutional project RVO 67985815.

This work was supported by TCHPC (Research IT, Trinity College Dublin). Calculations were performed on the Kelvin cluster maintained by the Trinity Centre for High Performance Computing. This cluster was funded through grants from the Higher Education Authority, through its PRTLI program. 

This research was partly funded by the Deutsche Forschungsgemeinschaft (DFG, German Research Foundation)
under Germany's Excellence Strategy - EXC-2094-390783311.

This research made use of \textsc{Tardis}, a community-developed software
package for spectral synthesis in supernovae \citep{Kerzendorf14,Kerzendorf19}. The development of \textsc{Tardis} received support from the Google Summer of Code initiative and from ESA's Summer of Code in Space program. \textsc{Tardis} makes extensive use of Astropy and PyNE. \textsc{Tardis} is a fiscally sponsored project of NumFOCUS.

\section*{Data Availability}

The data underlying this article are available on WISeREP, at
\url{https://wiserep.weizmann.ac.il/object/284} for ASASSN-14lp and \url{https://wiserep.weizmann.ac.il/object/251} for SN 2013dy.





\begin{thebibliography}{99}
\bibitem[\protect\citeauthoryear{Ashall et al.}{2014}]{Ashall14}
Ashall C., Mazzali P. A., Bersier D., Hachinger S., Phillips M., Percival S., James P., Maguire K. 2014, MNRAS, 445, 4424
\bibitem[\protect\citeauthoryear{Ashall et al.}{2016}]{Ashall16}
 Ashall C., Mazzali P. A., Pian E., James P. A., 2014, MNRAS, 463, 1891
\bibitem[\protect\citeauthoryear{Barna et al.}{2017}]{Barna17}
Barna B., Szalai T., Kromer M., Kerzendorf W., Vinkó J., Silverman J. M.,
Marion G. H., Wheelere J. C., 2017, MNRAS, 471, 4865
\bibitem[\protect\citeauthoryear{Barna et al.}{2018}]{Barna18}
Barna B., Szalai T., Kerzendorf W., Kromer M., Sim S. A., Magee M. R., Leibundgut B., 2018, MNRAS, 480, 3609
\bibitem[\protect\citeauthoryear{Baron et al.}{1996}]{Baron96}
Baron E., Hauschildt P. H., Nugent P., Branch D., 1996, MNRAS, 283, 297
\bibitem[\protect\citeauthoryear{Brown et al.}{2010}]{Brown10}
Brown P. J., Roming P. W. A., Milne P., et al., 2010, ApJ, 721, 1608
\bibitem[\protect\citeauthoryear{Brown et al.}{2015}]{Brown15}
Brown P. J., Baron E., Milne P., Roming P. W. A. and Wang L., 2015, ApJ, 809, 37
\bibitem[\protect\citeauthoryear{Boyle et al.}{2017}]{Boyle17}
Boyle A., Sim S. A., Hachinger S., Kerzendorf W., 2017, A\&A, 599, 46
\bibitem[\protect\citeauthoryear{Burns et al.}{2011}]{Burns11}
Burns C. R. et al., 2011, AJ, 141, 19
\bibitem[\protect\citeauthoryear{Casper et al.}{2013}]{Casper13}
Casper C. et al., 2013, Cent. Bur. Electron. Telegrams, 3588, 1
\bibitem[\protect\citeauthoryear{Chen et al.}{2019}]{Chen19}
Chen X., Hu L. \& Wang L., 2019, arXiv: 1911.05209
\bibitem[\protect\citeauthoryear{Dessart et al.}{2014}]{Dessart14}
Dessart L., Hillier J. D., Blondin S., Khokhlov A., 2014, MNRAS, 441, 3249
\bibitem[\protect\citeauthoryear{Fink, Hillebrandt \& R\"{o}pke}{2007}]{Fink07}
Fink M., Hillebrandt W. \& Röpke, F. K. 2007, A\&A, 476, 1133
\bibitem[\protect\citeauthoryear{Fink et al.}{2010}]{Fink10}
Fink M., Röpke F. K., Hillebrandt W., Seitenzahl I. R., Sim S. A. \& Kromer M., 2010, A\&A, 514, 53
\bibitem[\protect\citeauthoryear{Fink et al.}{2014}]{Fink14}
Fink M. et al., 2014, MNRAS, 438, 1762
\bibitem[\protect\citeauthoryear{Firth et al.}{2015}]{Firth15}
Firth R. E. et al., 2015, MNRAS, 446, 3895
\bibitem[\protect\citeauthoryear{Folatelli et al.}{2010}]{Folatelli10}
Folatelli G. et al. 2010, AJ, 139, 120
\bibitem[\protect\citeauthoryear{Foley et al.}{2016}]{Foley16}
Foley R. et al, 2016, MNRAS, 461, 1308
\bibitem[\protect\citeauthoryear{Gamezo et al.}{1999}]{Gamezo99}
Gamezo V. N., Wheeler J. C., Khokhlov A. M., Oran E. S., 1999, ApJ, 512, 827
\bibitem[\protect\citeauthoryear{Heringer et al.}{2019}]{Heringer19}
Heringer E., van Kerkwijk M. H., Sim S. A., Kerzendorf W. E., Graham Melissa L., 2019, ApJ, 871, 250
\bibitem[\protect\citeauthoryear{Hillebrandt et al.}{2013}]{Hillebrandt13}
Hillebrandt W., Kromer M., Röpke F. K., Ruiter A. J., 2013, Frontiers of Physics, 8, 116
\bibitem[\protect\citeauthoryear{Hoyle \& Fowler}{1960}]{Hoyle60}
Hoyle F., Fowler W. A., 1960, ApJ, 132, 565
\bibitem[\protect\citeauthoryear{Höflich et al.}{1998}]{Hoflich98}
Höflich P., Wheeler J. C. \& Thielemann F. K., 1998, ApJ, 495, 617
\bibitem[\protect\citeauthoryear{Höflich et al.}{2002}]{Hoflich02}
Höflich P., Gerardy C. L., Fesen R. A., Sakai S., 2002, ApJ 568, 791
\bibitem[\protect\citeauthoryear{Iben \& Tutukov}{1984}]{Iben84}
Iben I. Jr, Tutukov A. V., 1984, ApJS, 54, 335
\bibitem[\protect\citeauthoryear{Kerzendorf \& Sim}{2014}]{Kerzendorf14}
Kerzendorf W. E., Sim S. A., 2014, MNRAS, 440, 387
\bibitem[\protect\citeauthoryear{Kerzendorf et al.}{2019}]{Kerzendorf19}  
Kerzendorf W. E. et al., 2018, tardis-sn/tardis: TARDIS v2.0.2 release, Zenodo Software Release
\bibitem[\protect\citeauthoryear{Khokhlov}{1991}]{Khokhlov91}
Khokhlov A. M., 1991, A\&A, 245, 114
\bibitem[\protect\citeauthoryear{Kromer \& Sim}{2009}]{Kromer09}
Kromer M., Sim S. A., 2009, MNRAS, 398, 1809
\bibitem[\protect\citeauthoryear{Kromer}{2010}]{Kromer10}
Kromer M., Sim S. A., Fink M., R\"{o}pke F. K., Seitenzahl I. R., Hillebrandt W., 2010, ApJ, 719, 1067
\bibitem[\protect\citeauthoryear{Lentz et al.}{2000}]{Lentz00}
Lentz E. J., Baron E., Branch D., Hauschildt P. H. \& Nugent P. E., 2000, ApJ, 530, 966
\bibitem[\protect\citeauthoryear{Livne \& Arnett}{1995}]{Livne95}
Livne E. \& Arnett D. 1995, ApJ, 452, 62
\bibitem[\protect\citeauthoryear{Lucy}{2002}]{Lucy02}
Lucy L. B., 2002, A\&A, 384, 725
\bibitem[\protect\citeauthoryear{Lucy}{2003}]{Lucy03}
Lucy L. B., 2003, A\&A, 403, 261
\bibitem[\protect\citeauthoryear{Lucy}{2005}]{Lucy05}
Lucy, L. B. 2005, A\&A, 429, 19
\bibitem[\protect\citeauthoryear{Magee et al.}{2016}]{Magee16}
Magee M. R. et al., 2016, A\&A, 589, 89
\bibitem[\protect\citeauthoryear{Magee et al.}{2018}]{Magee18}
Magee M. R., Sim S. A., Kotak R., Kerzendorf W. E., 2018, A\&A, 614, 115
\bibitem[\protect\citeauthoryear{Magee et al.}{2020}]{Magee20}
Magee M. R., Maguire K., Kotak R., Sim S. A., Gillanders J. H., Prentice S. J., Skillen K., 2020, A\&A, 634, 37
\bibitem[\protect\citeauthoryear{Mazzali et al.}{2008}]{Mazzali08}
Mazzali P. A., Sauer D. N., Pastorello A., Benetti S., Hillebrandt W., 2008, MNRAS, 386, 1897
\bibitem[\protect\citeauthoryear{Mazzali et al.}{2014}]{Mazzali14}
Mazzali P. A. et al., 2014, MNRAS, 439, 1959
\bibitem[\protect\citeauthoryear{Nomoto et al.}{1984}]{Nomoto84}
Nomoto K., Thielemann F.-K., Yokoi K., 1984, ApJ, 286, 644
\bibitem[\protect\citeauthoryear{Pakmor et al.}{2011}]{Pakmor11}
Pakmor R., Hachinger S., Röpke F. K., Hillebrandt W., 2011, A\&A, 528, 117
\bibitem[\protect\citeauthoryear{Pan et al.}{2015}]{Pan15}
Pan Y.-C. et al., 2015, MNRAS, 452, 4307
\bibitem[\protect\citeauthoryear{Perlmutter et al.}{1999}]{Perlmutter99}
Perlmutter S. et al., 1999, ApJ, 517, 565
\bibitem[\protect\citeauthoryear{Phillips}{1993}]{Phillips93}
Phillips M. M., 1993, ApJ, 413, L105
\bibitem[\protect\citeauthoryear{Pinto \& Eastman}{2000}]{Pinto00}
Pinto P. A., Eastman R. G., 2000, ApJ, 530, 757
\bibitem[\protect\citeauthoryear{Piro \& Nakar}{2013}]{Piro13}
Piro A. L., Nakar E., 2013, ApJ, 769, 67
\bibitem[\protect\citeauthoryear{Piro \& Nakar}{2014}]{Piro14}
Piro A. L., Nakar E., 2014, ApJ, 784, 85
\bibitem[\protect\citeauthoryear{Piro \& Morozova}{2016}]{Piro16}
Piro A. L., Morozova V. S., 2016, ApJ, 826, 96
\bibitem[\protect\citeauthoryear{Polin et al.}{2019}]{Polin19}
Polin A., Nugent P., Kasen D., 2019, ApJ, 873, 84
\bibitem[\protect\citeauthoryear{Riess et al.}{1998}]{Riess98}
Riess A. G. et al., 1998, AJ, 116, 1009
\bibitem[\protect\citeauthoryear{Riess et al.}{2016}]{Riess16}
Riess A. G. et al., 2016, ApJ, 826, 56
\bibitem[\protect\citeauthoryear{R\"{o}pke et al.}{2012}]{Ropke12}
R\"{o}pke F. K. et al., 2012, ApJL, 750, 19
\bibitem[\protect\citeauthoryear{R\"{o}pke}{2017}]{Ropke17}
R\"{o}pke F. K., 2017, Handbook of Supernovae, Springer International Publishing AG, 1185
\bibitem[\protect\citeauthoryear{Sahu et al.}{2008}]{Sahu08}
Sahu D. K. et al., 2008, ApJ, 680, 580
\bibitem[\protect\citeauthoryear{Sauer et al.}{2008}]{Sauer08}
Sauer D. N., Mazzali P. A., Blondin S. et al., 2008, MNRAS, 391, 1605
\bibitem[\protect\citeauthoryear{Schlafly \& Finkbeiner}{2011}]{Schlafly11}
Schlafly E. F., Finkbeiner D. P., 2011, ApJ, 737, 103
\bibitem[\protect\citeauthoryear{Seitenzahl et al.}{2013}]{Seitenzahl13}
Seitenzahl I. R., 2013, MNRAS, 429, 1156
\bibitem[\protect\citeauthoryear{Seitenzahl et al.}{2016}]{Seitenzahl16}
Seitenzahl I. R. et al., 2016, A\&A, 592, 57
\bibitem[\protect\citeauthoryear{Shappee et al.}{2016}]{Shappee16}
Shappee B. J. et al., 2016, ApJ, 826, 144
\bibitem[\protect\citeauthoryear{Shen et al.}{2018}]{Shen18}
Shen K. J., Kasen D., Miles B. J., Townsley D. M., 2018, ApJ, 854, 52
\bibitem[\protect\citeauthoryear{Shen et al.}{2021}]{Shen21}
Shen K. J., Blondin S., Kasen D., Dessart L., Townsley D. M., Boos S., Hillier D. J., 2021, ApJ, 909, 18
\bibitem[\protect\citeauthoryear{Silverman et al.}{2015}]{Silverman15}
Silverman J. M., Vinkó J., Marion G. H., Wheeler J. C., Barna B., Szalai T., Mulligan B. W., Filippenko A. V., 2015, MNRAS, 451, 1973
\bibitem[\protect\citeauthoryear{Sim et al.}{2012}]{Sim12}
Sim S. A., Fink M., Kromer M.,Röpke F. K., Ruiter A. J. and Hillebrandt W., 2012, MNRAS, 420, 3003
\bibitem[\protect\citeauthoryear{Sim et al.}{2013}]{Sim13}
Sim S. A. et al., 2013, MNRAS, 436, 333
\bibitem[\protect\citeauthoryear{Stehle et al.}{2005}]{Stehle05}
Stehle M., Mazzali P. A., Benetti S., Hillebrandt W., 2005, MNRAS, 360,
1231
\bibitem[\protect\citeauthoryear{Timmes \& Woosley}{1992}]{Timmes92}
Timmes F. X., Woosley S. E., 1992, ApJ, 396, 649
\bibitem[\protect\citeauthoryear{Timmes et al.}{2003}]{Timmes03}
Timmes F. X., Brown E. F. \& Truran J. W. 2003, ApJL, 590, L83
\bibitem[\protect\citeauthoryear{Walker et al.}{2012}]{Walker12}
Walker E. S., Hachinger S., Mazzali P. A. et al., 2012, MNRAS, 427, 103
\bibitem[\protect\citeauthoryear{Webbink}{1984}]{Webbink84}
Webbink R. F., 1984, ApJ, 277, 355
\bibitem[\protect\citeauthoryear{Whelan \& Iben}{1973}]{Whelan73}
Whelan J., Iben I. Jr., 1973, ApJ, 186, 1007
\bibitem[\protect\citeauthoryear{Woosley \& Weaver}{1994}]{Woosley94}
Woosley S. E. \& Weaver T. A. 1994, ApJ, 423, 371
\bibitem[\protect\citeauthoryear{Woosley}{2007}]{Woosley07}
Woosley S. E., Kasen D., Blinnikov S., Sorokina E., 2007, ApJ, 662, 487
\bibitem[\protect\citeauthoryear{Woosley \& Kasen}{2011}]{Woosley11}
Woosley S. E., Kasen Daniel, 2011, ApJ, 734, 38
\bibitem[\protect\citeauthoryear{Zheng et al.}{2013}]{Zheng13}
Zheng W. et al., 2013, ApJ, 778, 15
\end{thebibliography}




\appendix
\label{appendix:a}

\section{Supplementary analysis: abundance tomography of SN 2013dy}
\label{appendix:sn2013dy}

SN 2013dy was discovered by the Lick Observatory Supernova Search program \citep{Casper13}. It appeared in the galaxy NGC 7250 at a redshift of $z = 0.00388$, whose distance was measured as $D=19.9$ Mpc via Cepheid variables \citep{Riess16}. By fitting the unfiltered band light curve with a broken power-law function, \cite{Zheng13} estimated the first-light time\footnote{We use the term ``time of first light'' as not equivalent to the ``time of explosion'', as the former refers to the moment when the first photons emerge, while the latter refers to the \textit{actual} moment, when the explosion starts. The intermediate ``dark phase'' can last for a few hours to days \citep{Piro13}.} of SN 2013dy to be MJD $56\,482.99$, only $2.4 \pm 1.2$ hours before the first detection, which makes the supernova one of the earliest detections of a SN Ia. \cite{Magee20} analyzed the light curve of SN 2013dy with the radiative transfer code TURTLS \citep[][for y brief description, see Sec. \ref{sec:turtls}]{Magee18} and constrained the date of explosion as MJD $56\,479.55 \pm 0.84$.

\begin{table*}
	\centering
	\caption{Log of the spectra of SN 2013dy; the phases are given with respect to B-band maximum. The times since explosion ($t_\rmn{exp}$) were fitting parameters for our TARDIS models (see in Sec. \ref{sec:fitting}) within $\pm$1.5 days of their estimated values in the referred papers.}
    \label{tab:log13dy}
    \begin{tabular}{cccccc}
		\hline
		MJD & Phase [days] & Telescope / Instrument & Wavelength [\r{A}] & Paper\\
		\hline
        \multicolumn{5}{c}{SN 2013dy}\\
		\hline
        56484.72 & $-16.4$ & Keck/DEIMOS & 4433 -- 9588 & \cite{Zheng13}\\
        56486.40 & $-14.7$ & Plaskett/Plaskett & 3649 -- 6944 & \cite{Zheng13}\\
        56488.42 & $-12.7$ & Lick/Kast & 3486 -- 10459 & \cite{Zheng13}\\
        56491.70 & $-9.4$ & FTN/FLOYDS-N & 3287 -- 10459 & \cite{Zheng13}\\
        56494.48 & $-6.6$ & HST/STIS & 1600 -- 10188 & \cite{Pan15}\\
        56498.60 & $-2.5$ & HST/STIS & 1600 -- 10188 & \cite{Pan15}\\
        56500.32 & $-0.8$ & HST/STIS & 1600 -- 10188 & \cite{Pan15}\\
        56502.31 & $+1.2$ & HST/STIS & 1600 -- 10188 & \cite{Pan15}\\
        56505.57 & $+4.5$ & HST/STIS & 1600 -- 10188 & \cite{Pan15}\\
        56509.49 & $+8.4$ & HST/STIS & 1600 -- 10188 & \cite{Pan15}\\
        \hline
        
        \hline
	\end{tabular}
\end{table*}

The Galactic component of the reddening is adopted from the dust emission map of \cite{Schlafly11} as $E(B-V)_\rmn{MW}=0.15$ mag, assuming $R_\rmn{V}=3.1$. \cite{Zheng13} estimated $E(B-V)_\rmn{host}=0.15$ mag based on the equivalent width (EW) measurement of the Na I D absorption. \cite{Pan15} used the light curve (LC) fitter SNooPy \citep{Burns11} and adopted $E(B-V)_\rmn{host}\,=\,0.20$ mag. Similarly to the case of ASASSN-14lp, the previously reported reddening values of SN 2013dy seem to be overestimated. In order to obtain better fits, we reduced the host reddening to $E(B-V)_\rmn{host}=0.07$ mag, which results in nearly the same goodness of fit for SNooPy. The Milky Way reddening is again adopted from the dust emission map of \cite{Schlafly11}, assuming $R_V\,=\,3.1$.

During the first thirty days after the explosion, SN 2013dy was observed by the Space Telescope Imaging Spectrograph (STIS) of the Hubble Space Telescope on six epochs. These spectra cover the mid-UV regime down to 1,600 \r{A} allowing critical information about the amount and distribution of IGEs to be obtained. However, the first HST spectrum has a phase of -6.6 days relative to B-band maximum, while constraining the properties of the outermost ejecta requires earlier epochs. Thus, we complemented the HST sample with four additional spectra from various sources, which allowed us to maintain the $\sim$3-day time resolution of the spectral series. The list of spectra used in our abundance tomography analysis can be found in Table \ref{tab:log13dy}.

\begin{figure*}
	\includegraphics[width=18.0cm]{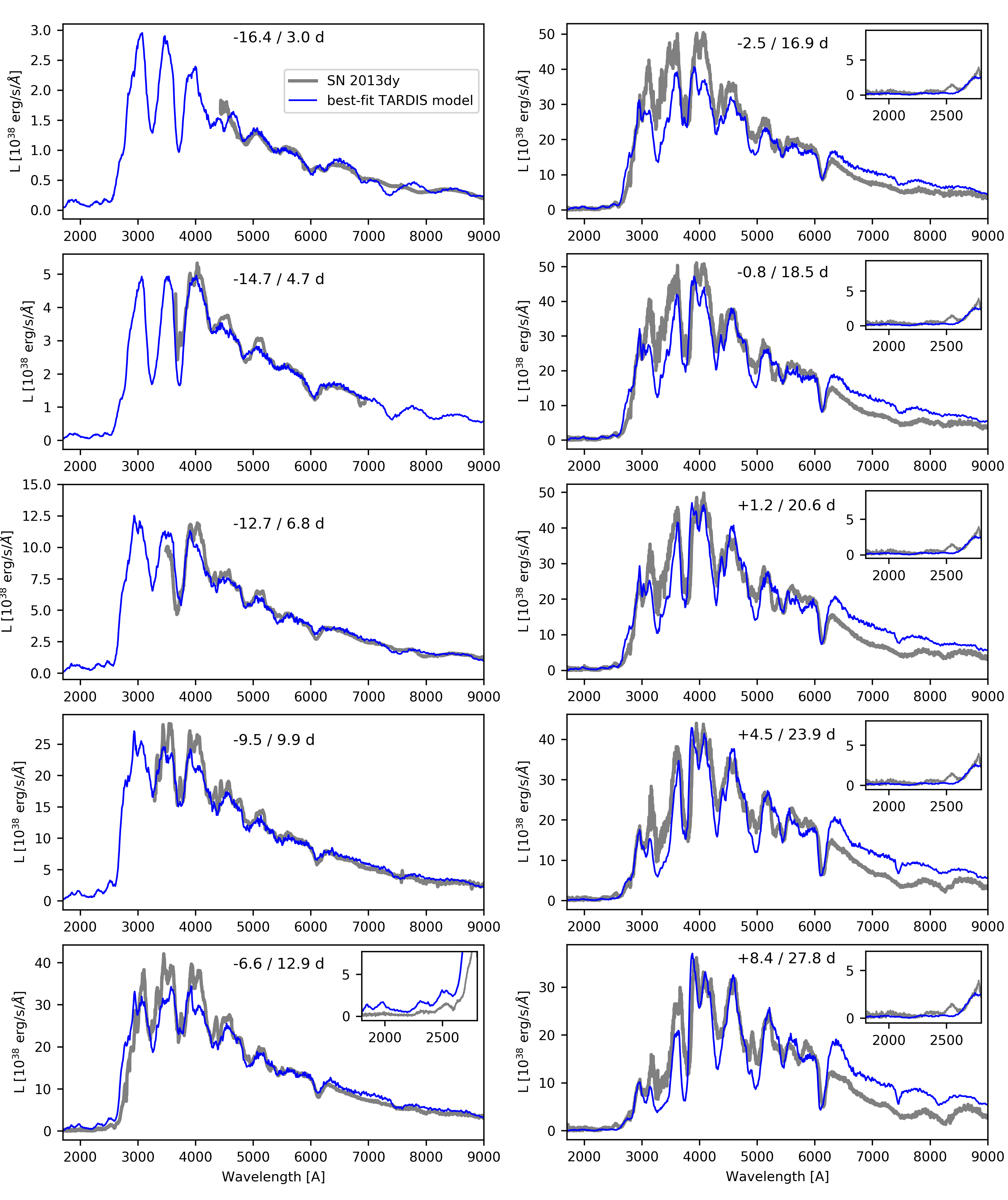}
    \caption{The studied spectral series of SN 2013dy (grey) and the corresponding best-fit TARDIS synthetic spectra (blue).}
    \label{fig:spectra_13dy}
\end{figure*}

\begin{figure*}
	\includegraphics[width=16cm]{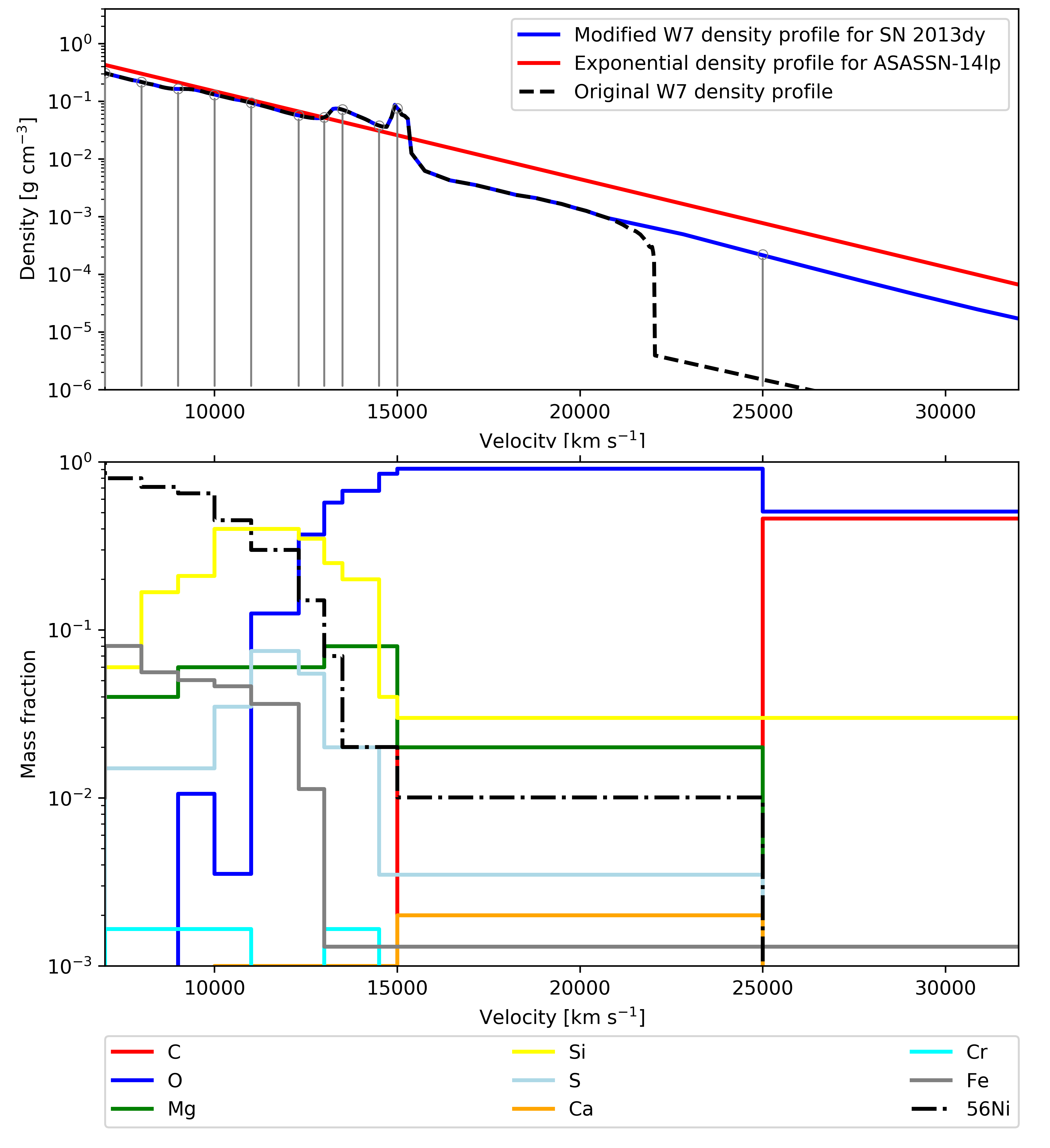}
    \caption{\textit{Top panel:} The modified density profile of W7 used for fitting the spectra of SN 2013dy (blue) and the best-fit pure exponential function for ASASSN-14lp (red) compared to the original W7 density structure at $t_\rmn{exp} = t_0 = 100$ s. The grey lines indicate the boundaries of the fitting layers for the corresponding chemical abundance models. \textit{Bottom panel:} The best-fit chemical abundance structure from the TARDIS fitting process of the Sn 2013dy spectral time series. The profile of the radioactive $^{56}$Ni shows the mass fraction at $t_\rmn{exp}=100$ s.}
    \label{fig:abundance_sn13dy}
\end{figure*}

\subsection{Spectral fitting}

For the model structure of SN 2013dy, we adopted the density profile of the W7 model \citep{Nomoto84} in order to reduce the number of fitting parameters. The very first test showed that such a structure can not reproduce the observed UV-suppression, because of the lack of material at high velocities. Thus, we modified the W7-profile by adding an exponential tail above 15\,000 km s$^{-1}$ and used it diluted with time for fitting all the spectra of SN 2013dy. Note that the extra mass from this tail is small and the new density profile is still consistent with $M_\rmn{Ch}$. The model ejecta was split into multiple layers, where the chemical abundances were fitting parameters to constrain the stratification of the chemical elements.

We set the boundaries of the fitting layers according to the photospheric velocity of each observed spectral epoch, similarly to the studies of \cite{Mazzali14} and \cite{Ashall14}. We fit the mass fractions of C, O, Mg, Si, S, Ca, Cr, Fe and $^{56}$Ni, also considering the decay products of the latter isotope. The mass fractions of every other chemical element heavier than He was set and fixed according to the solar metallicity.

When fitting each spectrum, we constructed the actual computation volume for the radiative transfer calculation based on the model structure described in the previous paragraph. The inner boundary of the computational volume is emitting pure blackbody radiation, thus, it is referred hereafter as photosphere, and its location as photospheric velocity ($v_\rmn{phot}$). The outer limit (i.e. the surface of the SN ejecta) of the volume is set to a high velocity, where the local density is sufficiently low for not having an impact on the computed spectra. According to our preliminary tests, we chose 35\,000 km s$^{-1}$ as the upper boundary in this study. The computational volume is divided into twenty radial layers, where layer boundaries are estimated according to equal mass regions. The mass fractions and the density value are sampled from the model structure and assumed to be uniform within a layer. The synthetic spectrum estimated by TARDIS based on the computational volume is compared to the observed spectrum and we modify the model structure to achieve a better match for the next TARDIS calculation. This iterative fitting allows us to constrain the physical and chemical properties that have a major impact on the output spectrum.

\subsection{Results from the abundance tomography}
\label{sec:results_13dy}

The final best-fit synthetic spectra of SN 2013dy can be found in Fig. \ref{fig:spectra_13dy}. There is a good match with the observed spectral series in general, although the fits fail to precisely reproduce the strong IGE absorption features between 3000 and 4500 \r{A} at some epochs. Note, however, that our study favours the fitting of the UV regime (see the insets of Fig. \ref{fig:spectra_13dy}) over the rest of the spectra.

The abundance tomography constrains the date of the explosion as MJD 56\,482.1, which is not in tension with the first detection at MJD 56\,483.09. Although the abundance tomography technique with TARDIS cannot provide a quantitative uncertainty about the date of explosion, the estimated value affects the fit of all the spectra in the analysis, especially the early epochs, which are most sensitive to the dilution of the assumed density profile. \cite{Zheng13} determine the time of first light to be almost a day later (MJD 56\,482.99 $\pm$ 0.05) based on the fit of the early light curve with a broken power-law function of $t^{1.80}$. Note that our analysis constrained the time of explosion, which is supposed to be earlier than the time inferred from a power-law fit of the light curve. \cite{Magee20} estimated MJD $56\,479.55 \pm 0.84$ for the same object. Thus, the time of explosion inferred from our fitting process seems consistent with the observations.

\begin{figure}
	\includegraphics[width=\columnwidth]{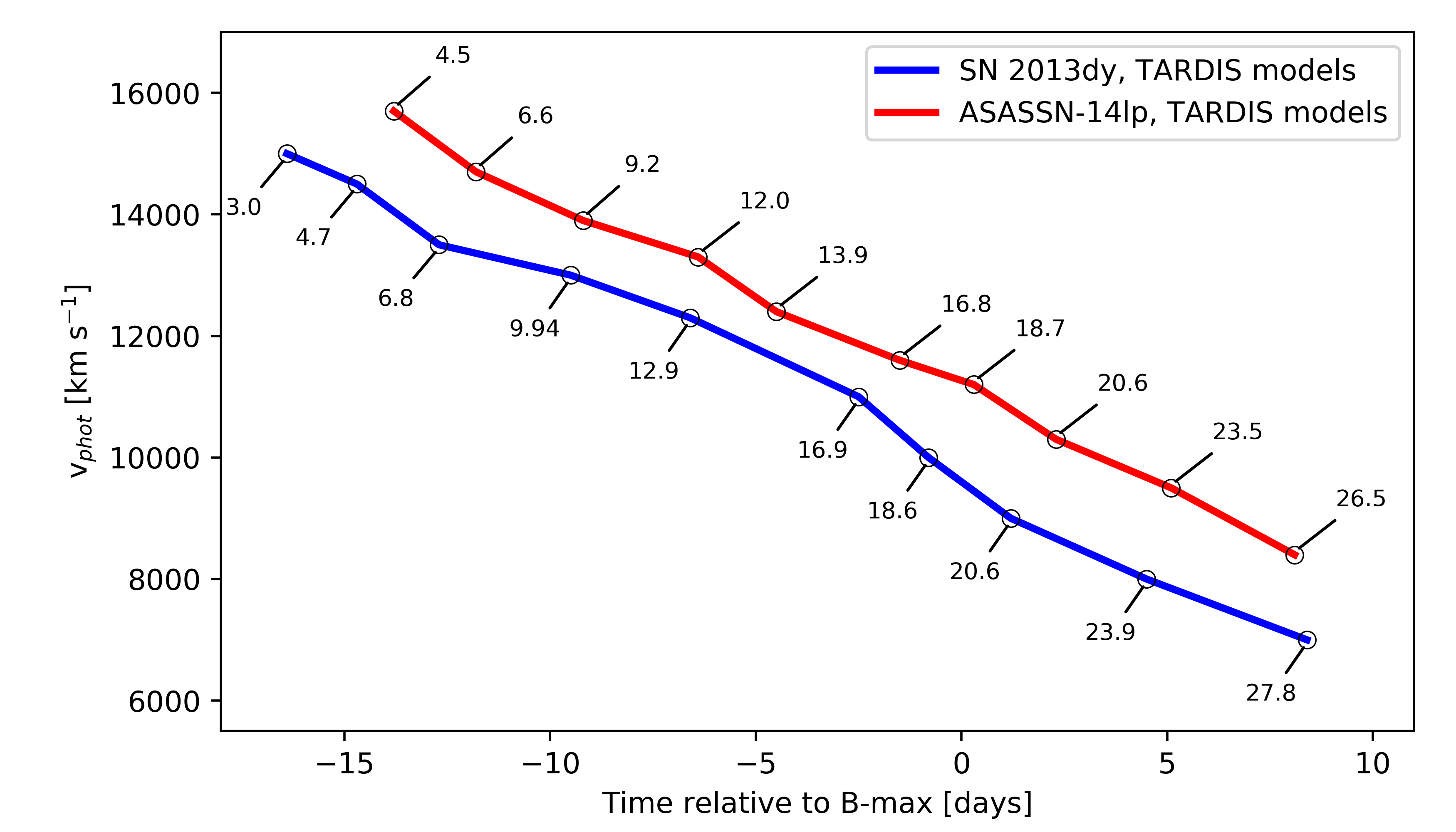}
    \caption{The velocity evolution of the photosphere in our modelling for SN 2013dy (blue) and ASASSN14-lp (red). The annotated numbers indicate the days since the best-fit explosion date.}
    \label{fig:velocities}
\end{figure}

The modified version of the W7 density profile can be seen in Fig. \ref{fig:abundance_sn13dy}. The chemical-abundance layers of the model are also shown in the plot. As described in Sec. \ref{sec:fitting}, the abundance layers are set according to the value of $v_\rmn{phot}$ for each observed epoch, except above 25\,000 km s$^{-1}$, where the abundances are assumed to be X(C) $\approx$ X(O) $\approx$ 0.50. This choice means that our modelling approach handles the whole region between 15\,000 (the $v_\rmn{phot}$ at the first epoch) and 25\,000 km s$^{-1}$ with uniform chemical abundances. The ejecta below 15\,000 km s $^{-1}$, by contrast, is well-sampled, with abundance layers of 500-700 km s$^{-1}$ width determined by the $v_\rmn{phot}$ at each spectral epoch. The evolution of $v_\rmn{phot}$ (Fig. \ref{fig:velocities}) is monotonically decreasing as it is expected. Although the decrease of $v_\rmn{phot}$ is not steady with time, the discrepancies are within the estimated $\sim 500$ km s$^{-1}$ accuracy of $v_\rmn{phot}$.

The inferred best-fit abundance profiles are presented in Fig. \ref{fig:abundance_sn13dy}. The outermost region above 25\,000 km s$^{-1}$ has no significant impact on the output synthetic spectra, thus, its mass fractions are fixed to almost exclusively unburnt material ($X(\rmn{O + C}) \approx 0.97$). The only exception is Si, which is present with $X(\rmn{Si}) = 0.03$ to correctly fit the blue wing of the Si II 
$\lambda6355$ line, probably affected by a high-velocity feature (HVF) in the earliest spectrum.

The next fitting region inwards, the layer between 15\,000 and 25\,000 km s$^{-1}$, has a key role in the origin of the mid-UV flux suppression (see below in Sec. \ref{sec:IGE_impact}). We find that a mass fraction of $^{56}$Ni of $X(^{56}\rmn{Ni})=0.01$ and its decay products at these high velocities can successfully reproduce the observed spectral features. In this outer region, oxygen remains the most abundant element with a mass fraction of 0.93. The high abundance of oxygen is not justified by the fitting of O I and O II features, but the lack of other chemical abundances (i.e. strong spectral lines requiring high mass fractions). Note that using oxygen as a `filler' element is a common choice in abundance tomography analysis \citep{Sahu08,Ashall14,Barna18}. Whether this dominant oxygen abundance is truly physical (i.e. oxygen from original WD matter or the product of carbon burning) or an artificial result of our fitting strategy (e.g. an incorrect choice of the density profile or "hidden" chemical elements) is uncertain.

Below 15\,000 km s$^{-1}$, the mass fraction of oxygen drops quickly, while the abundances of the IMEs and IGEs increase toward lower velocities. IMEs peak between 11\,000 and 13\,000 km s$^{-1}$, and Si is the most abundant element in this region with $X(\rmn{Si})=0.40$. Below 10,000 km s$^{-1}$, the mass fraction of Si starts to decrease and reaches $X(\rmn{Si})=0.06$ at 8\,000 km s$^{-1}$, the lower limit of our abundance tomography analysis. The abundance of sulfur, which peaks with $X(\rmn{S})=0.08$ at 11,000 km s$^{-1}$, follows the changes of the silicon mass fraction. The abundances of the other two IMEs that are subject to the fitting process, calcium and magnesium, are less well constrained. A low mass fraction of  $X(\rmn{Ca})=0.001-0.002$ is sufficient to reproduce the prominent Ca II H\&K feature at $\sim$3800 \r{A}. However, the same abundance causes only a weak absorption feature at the wavelengths of the Ca II NIR triplet. In the case of magnesium, there is no unblended feature at the observed wavelengths, but a few percents of Mg affect the fit positively, especially the red end of the mid-UV suppression.

Similarly to Sec. \ref{sec:IGE_impact}, we also tested the impact of IGE abundances at high velocities for the near-maximum spectrum of SN 2013dy. For this, we adopt the best-fit TARDIS model and vary the chemical composition in the outer fitting region (i.e. between 15\,000 and 25\,000 km s$^{-1}$). Note that the abundance tomography of SN 2013dy resulted in 1\% of $^{56}$Ni in this velocity region. Abundances in below and above this region remained the same as in our best-fit TARDIS model. The comprehensive tests support the conclusion, that the UV suppression can be reproduced either by assuming extremely high progenitor metallicity or by assuming an extended $^{56}$Ni distribution.

\begin{figure}
	\includegraphics[width=\columnwidth]{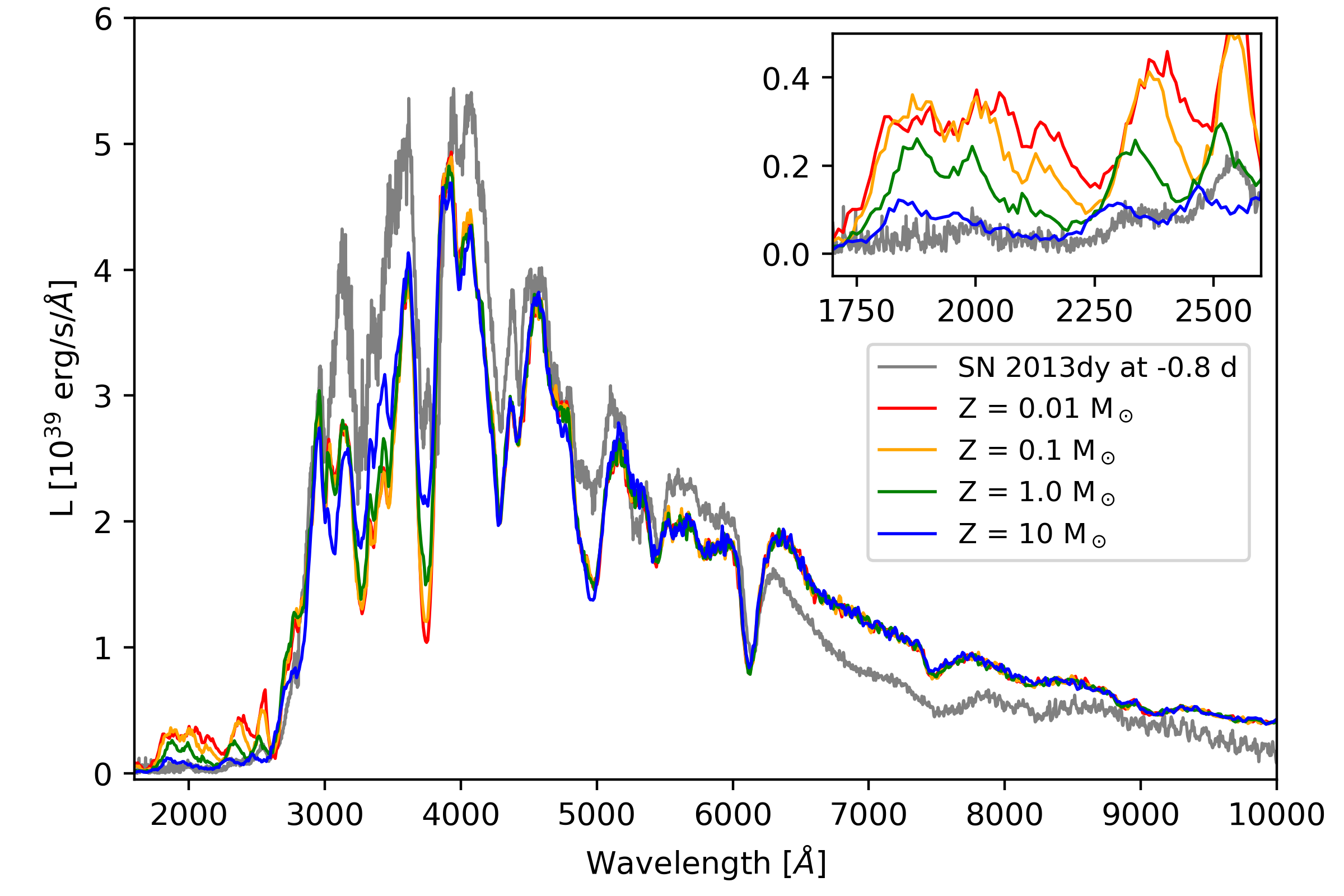}
    \caption{Model spectra with $X(^{56}\rmn{Ni}) = 0$ and different progenitor metallicities in the outermost fitting layer, i.e. between 15\,000 and 25\,000 km s$^{-1}$ in the model structure. The spectrum of SN 2013dy obtained at -0.8 day is also plotted (grey).}
    \label{fig:metallicity_test_13dy}
\end{figure}

\begin{figure}
	\includegraphics[width=\columnwidth]{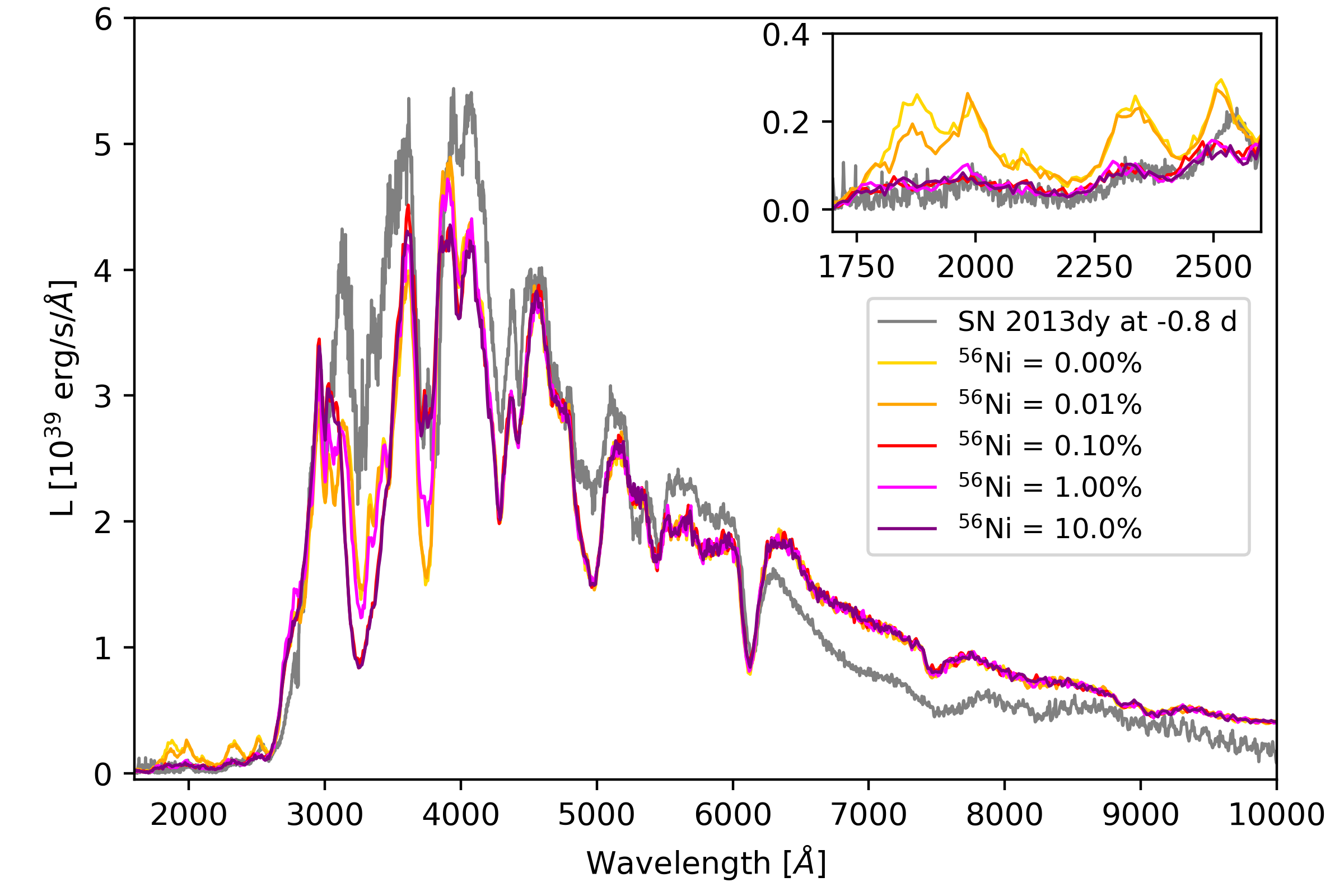}
    \caption{Model spectra with the same progenitor metallicity ($Z=Z_\odot$) and different mass fractions of $^{56}$Ni in the outermost fitting layer, i.e. between 15\,000 and 25\,000 km s$^{-1}$ in the model structure. The spectrum of SN 2013dy obtained at -0.8 day is also plotted (grey).}
    \label{fig:nickel_test_13dy}
\end{figure}


\bsp	
\label{lastpage}
\end{document}